\documentclass[aps,pre,epsf,superscriptaddress,amsmath,amssymb,amsfonts,twocolumn,showpacs]{revtex4-1}

\usepackage{graphicx}
\usepackage{epsfig}
\usepackage{dcolumn}
\usepackage{bm}
\usepackage{braket}
\usepackage{amsmath}
\usepackage{color}
\newcommand{\abs}[1]{\left| #1 \right|} 
\usepackage{hyperref}
\newcommand*\diff{\mathop{}\!\mathrm{d}} 
\DeclareMathOperator{\sech}{sech} 

\usepackage{makecell}

\begin{document}

\title{Controlled Generation of Dark-Bright Soliton Complexes \\
in Two-Component and Spinor Bose-Einstein Condensates }

\author{A. Romero-Ros}
\affiliation{Center for Optical Quantum Technologies,
Department of Physics,
University of Hamburg,
Luruper Chaussee 149,
22761 Hamburg,
Germany}

\author{G. C. Katsimiga}
\affiliation{Center for Optical Quantum Technologies,
Department of Physics,
University of Hamburg,
Luruper Chaussee 149,
22761 Hamburg,
Germany}

\author{P. G. Kevrekidis}
\affiliation{Department of Mathematics and Statistics, University
of Massachusetts Amherst, Amherst, MA 01003-4515, USA}

\author{P. Schmelcher}
\affiliation{Center for Optical Quantum Technologies,
Department of Physics, University of Hamburg,
Luruper Chaussee 149,
22761 Hamburg,
Germany}
\affiliation{The Hamburg Centre for Ultrafast Imaging,
University of Hamburg, Luruper Chaussee 149, 22761 Hamburg,
Germany}

\date{\today}

\begin{abstract}

We report on the controlled creation of multiple soliton complexes of the dark-bright type in one-dimensional
two-component, three-component and spinor Bose-Einstein condensates. 
The formation of solitonic entities of the dark-bright type 
is based on the so-called matter wave interference of spatially separated condensates.
In all three cases a systematic numerical study is carried out upon considering different variations of each 
systems' parameters both in the absence and in the presence of a harmonic trap.
It is found that manipulating the initial separation or the chemical 
potential of the participating components allows us to tailor the number of nucleated dark-bright states.  
Particularly, the number of solitons generated increases upon increasing either the initial separation 
or the chemical potential of the participating components.
Similarities and differences of the distinct models considered herein are showcased, 
while the robustness of the emerging states is illustrated via direct
numerical integration demonstrating their long time propagation.
Importantly, for the spinorial system, we unravel the existence of beating dark soliton arrays that 
are formed due to the spin-mixing dynamics. 
These states persist in the presence of a parabolic trap, often relevant for associated
experimental realizations.
\end{abstract}

\maketitle

\section{Introduction} \label{sec:introdcution}

Among the nonlinear excitations that arise in Bose-Einstein condensates
(BECs)~\cite{Anderson1995, Davis1995}, 
matter-wave dark~\cite{Frantzeskakis_2010} and bright~\cite{tomio}
solitons constitute the fundamental signatures. 
These structures stem from the balance between dispersion and nonlinearity and exist in single component BECs with 
repulsive and attractive interparticle interactions respectively~\cite{Zakharov1972,Zakharov1973}.
Also more complex structures consisting of dark solitons in one component and bright solitons hosted in the second component of 
a binary BEC have been experimentally realized~\cite{Becker2008,Middelkamp2011,Hamner2011,Yan2011,Hoefer2011,Yan2012}. 
The existence and robustness of a single dark-bright (DB) soliton as well as 
interactions between multiple DB states both with each other as well as with 
impurities have been exhaustively studied in such 
settings~\cite{brazhnyi2011stable,yin2011coherent,Yan2011,Achilleos2011,Alvarez2013,yan2015dark,karamatskos2015stability,
katsimiga2017dark,katsimiga2017stability,katsimiga2018dark}.
In contrast to single component setups, DB solitons are the building blocks that emerge in repulsively interacting 
two-component BECs~\cite{Kevrekidis2016}. In such a repulsive environment
(where bright solitonic states cannot exist on their own)
DB states owe their existence to the effective potential 
created by each of the participating dark solitons 
into which each of the bright solitons is trapped and 
consequently waveguided~\cite{Trillo1988,Christodoulides1988,Ostrovskaya1998}. 
This waveguiding notion has been firstly introduced in the context of nonlinear 
optics~\cite{Afanasyev1989,Kivshar1993,Christodoulides1996,Buryak1996,Sheppard1997,Chen1997,Kivshar1998,Ostrovskaya1999,Park2000}.  
Besides the aforementioned two-component BECs, the experimental realization of 
spinor BECs~\cite{Stamper-Kurn2001,Chang2004,Chang2005,Kawaguchi2012,Stamper-Kurn2013}
offers 
new possibilities of investigating the different soliton entities that arise in 
them~\cite{Ohmi1998,Tsuchida1998,Ieda2004,Ieda2004a,Ieda2005,Li2005,Wadati2005,Ieda2006,Uchiyama2006,Ieda2007,Zhang2007,
Kurosaki2007,Dabrowska-Wuster2007,Kawaguchi2012,Stamper-Kurn2013}.
In this context, more complex compounds in the form of dark-dark-bright (DDB) and dark-bright-bright (DBB)
solitons have been theoretically predicted~\cite{Nistazakis2008,Xiong2010} 
and very recently experimentally observed~\cite{Bersano2018}.

There are multiple ways of generating single and multiple dark solitons~\cite{katsimiga2018many}
(with the latter sometimes referred to as the dark soliton train~\cite{Brazhnyi2003}), 
in single component BECs. 
Common techniques consist of density engineering~\cite{Dutton2001,Engels2007,Shomroni2009}, 
phase engineering~\cite{Burger1999,Denschlag2000,Anderson2001,Becker2008}, 
and collision of two spatially separated condensates~\cite{Weller2008,hoefer2009matter}
(see also~\cite{bpa} for an interesting geometric higher dimensional
implementation of the latter process so as to produce vortices).
This latter generation process can be thought of as a consequence of matter
wave interference 
of the two condensates~\cite{Reinhardt1997,Scott1998,Weller2008,Theocharis2010}.
Additionally, also known are the conditions under which the controllable formation of dark soliton trains can be 
achieved~\cite{Reinhardt1997,Scott1998,Brazhnyi2003,Weller2008,Theocharis2010}.
In particular, it has been demonstrated that the number of generated dark solitons depends on the phase and momentum of 
the colliding condensates~\cite{Weller2008,Theocharis2010}. 
On the contrary, in multi-component settings such as two-component and spinor BECs the dynamics is much more involved. 
In this context, large scale counterflow experiments exist according to  
which also DB soliton trains can be created~\cite{Hamner2011}. 
However, to the best of our knowledge a systematic study regarding the controllable formation of these more complex solitonic 
structures and their relevant extensions in spinorial BECs is absent in the current literature.
This controlled formation process represents the core of the present investigation.

Motivated by recent experimental advances in one-dimensional (1D) 
two-component~\cite{Hamner2011,Middelkamp2011,Yan2011,Hoefer2011,Yan2012} and more importantly spinor 
BECs~\cite{Bersano2018}, here we report on the controllable generation of multiple soliton complexes. 
These include DB solitons in two-component BECs, 
and variants of these structures, i.e. DDB and DBB soliton arrays,
in three-component and spinor BECs.
For all models under consideration, 
the creation process of the relevant states is based on the so-called matter wave interference 
of separated condensates being generalized to multi-component systems. 
In all cases, the homogeneous setting is initially discussed 
and subsequently we generalize our findings to the case where an experimentally
motivated parabolic confinement, i.e. trap, is present.

Specifically, for the homogeneous settings investigated herein 
the creation process is as follows. 
To set up the interference dynamics,
an initial inverted rectangular pulse (IRP) 
is considered~\cite{Zakharov1973} for the component(s) that will 
host later on the dark solitons. 
The counterflow process relies on the collision of the two sides of the
pulse. 
For the remaining component(s), that will 
host later on the bright solitons, a Gaussian pulse initial condition
is introduced.
It is shown that such a process ensures the formation of dark soliton arrays
the number of solitons in
which can be manipulated by properly adjusting the width 
of the initial IRP.
Additionally, the dispersive behavior of the Gaussian used, 
due to the defocusing nature of each system, 
allows its confinement in the effective potential created by each of the
formed dark solitons 
and thus leads to the formation of the desired localized humps.
The latter are trapped and subsequently waveguided by the corresponding dark solitons.
In this way, arrays of robustly propagating DB, DDB and DBB solitons
in two-component, three-component and spinor systems are showcased.
Indeed, and as far as the two-component system is concerned, 
we verify among others, that during evolution the trajectory of each of the nucleated pairs 
follows the analytical predictions stemming from the exact single DB state.
Also for the three-component scenario generalized expressions for the soliton characteristics are 
extracted and deviations from the latter when different initializations are considered are
discussed in detail.  
In the spinor setting the controlled nucleation
of arrays consisting of multiple DBB and DDB solitons is demonstrated,
a result that can be tested in current state-of-the-art experiments~\cite{Bersano2018}.
Remarkably enough, in the DDB nucleation process, 
the originally formed DDB arrays soon after their formation transition
into beating dark solitons that gradually arise in all three hyperfine 
components~\cite{Park2000,Ohberg2001,Hoefer2011,Yan2012}.  
This transition stems, as we will explain in more detail below,
from the spin-mixing dynamics that allows for particle 
exchange among the hyperfine components.

After the proof-of-principle in the spatially homogeneous case, 
we turn to the harmonically trapped models, where once
again in order to induce the dynamics, 
counterflow
techniques are utilized~\cite{Weller2008,Theocharis2010,Hamner2011}.
Now the background on top of which the spatially separated BECs are initially set
up asymptotes to a 
Thomas-Fermi (TF) profile for all the participating components.
The counterflowing components are initially relaxed
in a double-well potential, while the other
component encounters a tight harmonic trap. 
The system is then released and evolves in a common
parabolic potential.  
It is found that by properly adjusting the initial separation of the
condensates 
or the chemical potential in each of the participating components leads to   
the controlled nucleation of a desired number of soliton structures, in this
case too, with similar functional dependences of the soliton number
on the system characteristics as above. 
For the two- and three-component systems it is found that
the generated soliton arrays travel within the parabolic trap 
oscillating and interacting with one another for large evolution times.
Finally, in the genuine spinor case and for a DDB formation process
again arrays but of oscillating and interacting beating dark solitons 
emerge in all hyperfine components. 
We find that these states
occur earlier in time when compared to the homogeneous scenario.
The spin-mixing dynamics is explained
via monitoring the population of the three hyperfine states.
Damping oscillations of the latter are observed in line
with the predictions in spinor $F=1$ BECs~\cite{Pu1999,Chang2005}. 

The work-flow of this presentation proceeds as follows. 
In Sec.~\ref{sec:model_setup} we present the different models under consideration. 
In particular, the spinor $F=1$ BEC system is initially introduced 
and the complexity of the model is reduced
all the way down to the single-component setting. 
Subsequently,  a brief discussion summarizing prior 
results regarding the controllable generation of dark soliton trains 
emerging in single-component systems is provided. 
Finally here, we comment on the initial state preparation utilized herein in order 
to controllably generate multiple soliton complexes of the DB type in multi-component BECs. 
Sec.~\ref{sec:results} contains our numerical findings ranging from two-component to spinor BEC systems.
In all the cases presented, the homogeneous setting is initially investigated, 
and we next elaborate on the relevant findings in the presence of traps. 
To conclude this work, in Sec.~\ref{sec:summ_concl} we summarize our findings  
and we also discuss future directions.


\section{Models and setups} \label{sec:model_setup}

\subsection{Equations of motion} \label{subsec:models}
We consider a 1D harmonically confined spinor $F=1$ BEC.
Such a system can be described by three coupled Gross-Pitaevskii equations (CGPEs),
one for each of the three hyperfine states $m_F=-1,0,+1$, of e.g. a $^{87}$Rb gas.   
In the mean-field framework the wavefunctions, ${\bf \Psi}(x,t)=\left[\Psi_{+1}(x,t),\Psi_{0}(x,t),\Psi_{-1}(x,t)\right]^T$, 
of the aforementioned hyperfine components are known to obey the following 
GPEs (see e.g.~\cite{Stamper-Kurn2013,Bersano2018}): 
\begin{subequations}\label{eq:spinor_hamiltonian}
\begin{eqnarray}
i\partial_t\Psi_{\pm 1}&= {\cal{H}}_0\Psi_{\pm 1}
+ g_n \left( |\Psi_{+1}|^2+ |\Psi_0|^2+|\Psi_{-1}|^2 \right) \Psi_{\pm1}  \nonumber \\
&+g_s\left(|\Psi_{\pm 1}|^2+|\Psi_{0}|^2-|\Psi_{\mp 1}|^2\right)\Psi_{\pm 1}  
+g_s\Psi_0^2\Psi^*_{\mp 1}, \nonumber \\ 
\label{eq:spinor_hamiltonian_a}
\end{eqnarray}
\begin{eqnarray}
i\partial_t\Psi_{0}&= {\cal{H}}_0\Psi_{0}
+g_n\left(|\Psi_{+1}|^2+|\Psi_0|^2+|\Psi_{-1}|^2\right)\Psi_{0} \nonumber \\
&+g_s\left(|\Psi_{+1}|^2+|\Psi_{-1}|^2\right)\Psi_0  
+2g_s\Psi_{+1}\Psi^*_{0}\Psi_{-1}. \nonumber \\
\label{eq:spinor_hamiltonian_b}
\end{eqnarray} 
\end{subequations}
In the above expressions the asterisk denotes the complex conjugate and 
${\mathcal{H}}_0=-\frac{1}{2}\partial^2_x+V(x)$ is the single-particle Hamiltonian. 
Here, $V(x)=\left(1/2\right)\Omega^2 x^2$ denotes  (unless indicated otherwise)
the external harmonic potential with frequency $\Omega=\omega_x/\omega_{\perp}$ and $\omega_{\perp}$
is the trapping frequency in the transverse direction.
Eqs.~(\ref{eq:spinor_hamiltonian_a})-(\ref{eq:spinor_hamiltonian_b}) were made dimensionless
by measuring length, time, and energy in units: $a_{\perp}=\sqrt{\hbar/(M\omega_{\perp})}$, 
$\omega_{\perp}^{-1}$, and $\hbar \omega_{\perp}$ respectively.
Here, $a_{\perp}$ is the transverse oscillator length.
In this work we consider condensates consisting of $^{87}$Rb atoms of mass $M$, 
and we assume strongly anisotropic clouds having a transverse trapping frequency 
$\omega_{\perp}=2\pi \times 175$~Hz $\gg \omega_x$ that is typically used in 
experiments with spinor $F=1$ BECs of $^{87}$Rb atoms~\citep{Bersano2018}.

In general, spinor BECs exhibit both symmetric or spin-independent and 
asymmetric or spin-dependent interatomic interactions.
In particular, $g_n$ is the so-called spin-independent interaction strength
being positive (negative) for repulsive (attractive) interatomic interactions.
$g_s$ denotes the so-called spin-dependent interaction strength  
being in turn positive (negative) for antiferromagnetic (ferromagnetic) interactions~\cite{Ho1998}.
Specifically, for a 1D spin-1 BEC $g_n=\frac{2(a_0+2a_2)}{3a_{\perp}}$ and $g_s=\frac{2(a_2-a_0)}{3a_{\perp}}$.
Here, $a_0$ and $a_2$ are the corresponding $s$-wave scattering lengths of two atoms in the scattering channels  
with total spin $F=0$ and $F=2$, respectively. 
The measured values of the aforementioned scattering lengths for $^{87}$Rb are
$a_0=101.8a_B$ and $a_2=100.4a_B$ where $a_B$ is the Bohr radius, 
resulting in a ferromagnetic spinor BEC~\cite{Klausen2001,vanKempen2002}. 

Finally, the total number of particles and the total magnetization for the 
system of Eqs.~(\ref{eq:spinor_hamiltonian_a})-(\ref{eq:spinor_hamiltonian_b}) are defined as
$N=\sum_{m_F} \int |\Psi_{m_F}|^2 \text{d}x$, and $M_z=\int \left(|\Psi_{+1}|^2-|\Psi_{-1}|^2\right) \text{d}x$, 
respectively.

Simplified BEC models can be easily obtained from Eqs.~(\ref{eq:spinor_hamiltonian_a})-(\ref{eq:spinor_hamiltonian_b}).
In particular, when the spin degrees of freedom are frozen, namely for $g_s=0$,
the aforementioned system reduces to the following three-component one 
\begin{equation}
i\partial_t\Psi_j={\mathcal{H}}_0\Psi_j+g_n\left(|\Psi_j|^2+|\Psi_k|^2+|\Psi_l|^2\right)\Psi_j.
\label{eq:3-CGPE}
\end{equation}
The indices $j,k,l$ here refer to each of the three $m_F=+1,0,-1$ components,
with $j \neq k \neq l$.
This three-component system, in the absence of an external confinement (i.e.,
for $V(x)=0$) and for constant $g_n$ which, without loss of 
generality, can be set to $g_n=1$,  
is said to be integrable and reduces to the so-called Manakov model~\cite{Tsuchida1998,Ieda2007,Manakov1973}. 
As such it admits exact soliton solutions of the DDB and DBB type~\cite{biondini2016three}.
Accounting for repulsive inter- and intra-species interactions (up
to a rescaling), we will set $g_n=1$ in our subsequent results discussion. 
Additionally, the two-component BEC can be retrieved by setting e.g. $\Psi_l=0$ in 
Eq.~(\ref{eq:3-CGPE}). 
Note that such a binary mixture consists of two different spin states, 
e.g. one with $\ket{F=1}$ and one with
$\ket{F=2}$, of the same atomic species and is
theoretically described by the following 
GPEs~\cite{kevrekidis2015defocusing}
\begin{equation}
i\partial_t\Psi_j=\mathcal{H}_0 \Psi_j + g_n \left(|\Psi_j|^2 + |\Psi_k|^2 \right) \Psi_j.
\label{eq:CGPE}
\end{equation}
Here, the indices $j,k$ refer to each of the two participating species.
Finally, the single-component case is retrieved by setting in Eq.~(\ref{eq:CGPE}) $\Psi_k=0$.
The corresponding GPE reads~\cite{Gross1961,Pitaevskii1961} 
\begin{equation}
i\partial_t\Psi=\mathcal{H}_0 \Psi + g_n |\Psi|^2\Psi.
\label{eq:GPE}
\end{equation}
In the forthcoming section we will first focus on the integrable version of Eq.~(\ref{eq:GPE})
and the exact arrays of dark soliton solutions that it admits.

\subsection{Prior analytical considerations and initial state preparation} \label{subsec:setup_homo}

It is well-known and experimentally confirmed that multiple dark solitons
can be systematically generated in single-component BECs, 
via the so-called matter wave interference of two initially separated 
condensates~\cite{Reinhardt1997,Scott1998,Weller2008,Theocharis2010}.  
Aiming to generalize this mechanism to multi-component systems, below we briefly discuss
previous studies on this topic. 

In particular, the problem of determining the parameters of a dark soliton formed by 
an initial excitation on a uniform background has been analytically
solved by the inverse scattering method~\cite{Zakharov1973}. 
In this framework, Eq.~(\ref{eq:GPE}) (with $V(x)=0$) is associated 
with the Zakharov-Shabat (ZS)~\cite{Zakharov1973, Espinola-Rocha2009} linear spectral problem.
The corresponding soliton parameters are related to the eigenvalues of this spectral problem,
calculated for a given initial condensate wavefunction $\Psi(x,0)$. 
Specifically, let us assume that $\Psi(x,0)$ has the form corresponding to an IRP
\begin{eqnarray}
\Psi(x,0)&=&u_0 \hspace{1.3cm} \text{at} \hspace{0.67cm} x<-a \nonumber,
\\
\Psi(x,0)&=&0 \hspace{1.5cm} \text{at} \hspace{0.5cm} -a<x<a \nonumber,
\\
\Psi(x,0)&=&u_1 e^{i\Delta\phi} \hspace{0.65cm} \text{at} \hspace{0.67cm} x>a,
\label{eq:square_well}
\end{eqnarray}
with $a$, $u_0$, $u_1$, $\Delta\phi$ denoting respectively the half-width, the
two amplitudes
and the phase difference between the two sides of the IRP.
Subsequently, for the case of $|u_0|=|u_1|=|u|$, it has been shown~\cite{Zakharov1973}
that the number of dark soliton pairs depends on the amplitude, $|u|$, 
and the phase difference $\Delta\phi$ of the initial IRP.
Namely, for $\Delta\phi=0$ which corresponds to a symmetric or in-phase (IP) IRP 
[see Eq.~(\ref{eq:square_well})],  
there exist $n$-symmetrical pairs of dark soliton solutions that are given by the solutions
of the (transcendental) eigenvalue equations
\begin{equation}
\abs{u}\cos(2a_n\lambda_n)=\pm\lambda_n.
\label{eq:cos}
\end{equation}
Here, $\lambda_n$, are the corresponding eigenvalues being bounded within the interval $[0,~|u|]$.
Importantly, solutions of Eq. (\ref{eq:cos}) exist only within the intervals 
$2a_n\lambda_n \in \left[(n-1)\pi,\left(n-\frac{1}{2}\right)\pi\right]$ with $n=1,2,3,\dots$.
Notice also, that for $n=1$ Eq. (\ref{eq:cos}) has at least one root within the interval 
$0< 2a_n\lambda_n <\frac{\pi}{2}$. Thus, there exists at least $1$-pair of coherent structures. 
Multiple roots of Eq. (\ref{eq:cos}) can be found but for appropriate 
values of the half-width $a_n$ that lie within the aforementioned interval. 
Therefore, there exists a threshold for the width $a_n$ above which solitons can be created.  
It has been demonstrated that the  
lower bound for the width of the IP-IRP in order to obtain $n$-symmetrical pairs of soliton solutions 
has the form
\begin{equation}
W_{IP}=2a_n=\frac{(n-1)\pi}{|u|}.
\label{eq:w_n_IP}
\end{equation}
In the above expression we have defined $W_{IP}\equiv 2a_n$.
Moreover, as dictated by Eq.~(\ref{eq:cos}) the total number of solitons is always even. 
Additionally, in order to obtain at least $1$-pair of soliton solutions, i.e. for $n=1$, then $W_{IP}> 0$
according to Eq.~(\ref{eq:w_n_IP}).

On the other hand, for $\Delta\phi=\pi$ [see Eq.~(\ref{eq:square_well})],
i.e. for an asymmetric IRP or out-of-phase (OP) initial conditions, the $n$-pairs of soliton solutions are given by
the following eigenvalue equations
\begin{equation}
\abs{u}\sin(2a_n\lambda_n)=\pm\lambda_n.
\label{eq:sin}
\end{equation}
Here, $2a_n\lambda_n \in \left[\left(n-\frac{1}{2}\right)\pi, n\pi\right]$ with $n=1,2,3,\dots$.
In the OP case the corresponding threshold for the width $2a_n$ reads [see Eq.~(\ref{eq:sin})] 
\begin{equation}
W_{OP}=2a_n=\frac{(n-\frac{1}{2})\pi}{|u|},
\label{eq:w_n_OP}
\end{equation}
where $W_{OP}\equiv 2a_n$ is introduced. 
For both IP- and OP-IRPs the amplitude, $\nu_n$, 
of each dark soliton pair is defined by the eigenvalues $0\leq |\lambda_n| \leq |u|$
through the relation $\nu_n=\sqrt{|u|^2-\lambda_n^2}$. 
Also each soliton's velocity is given by $v_n=\pm\lambda_n$.
From Eq.~(\ref{eq:w_n_OP}) and for $n=1$, 
we can again obtain the minimum width to assure the existence of at least a 1-pair solution.
The latter reads $W_{OP}=\pi/(2|u|)$. 
Although Eq.~(\ref{eq:sin}) gives the solutions for $n$-pairs of solitons, there exists also 
an isolated wave for $\lambda=0$ corresponding to 
a black soliton with $\nu=|u|$ and $v=0$.
Summarizing, an odd number of dark solitons 
is expected to be generated for OP initial conditions.
We should remark at this point that Eq.~(\ref{eq:w_n_IP}) and Eq.~(\ref{eq:w_n_OP})
dictate the dependence of the generated number of dark solitons not only on the phase,
but also on the momenta (through the relation $v_n=\pm\lambda_n$) 
of the colliding condensates~\cite{Weller2008,Theocharis2010}.  
In particular, for larger initial widths 
the number of dark solitons generated increases
since the two sides of the IRP acquire, during the counterflow, larger momenta 
(see also here the works of Refs.~\cite{Ostrovskaya1999,Nikolov2004} and references 
therein for relevant studies in nonlinear optics).
Importantly, also, the effective intuition
of the number of solitons filling in the space between the two sides
in ``units'' of the healing length, namely in dark solitons, is
a relevant one to qualitatively bear in mind.
Finally, we must also note that in the BEC context an initial state preparation
having the form of Eq.~(\ref{eq:square_well}) can, in principle, be achieved 
by standard phase imprinting methods and the
use of phase masks~\cite{Denschlag2000,scherer2007vortex,Becker2008}.

\begin{figure}[t]
\centering
\includegraphics[width=0.48\textwidth]{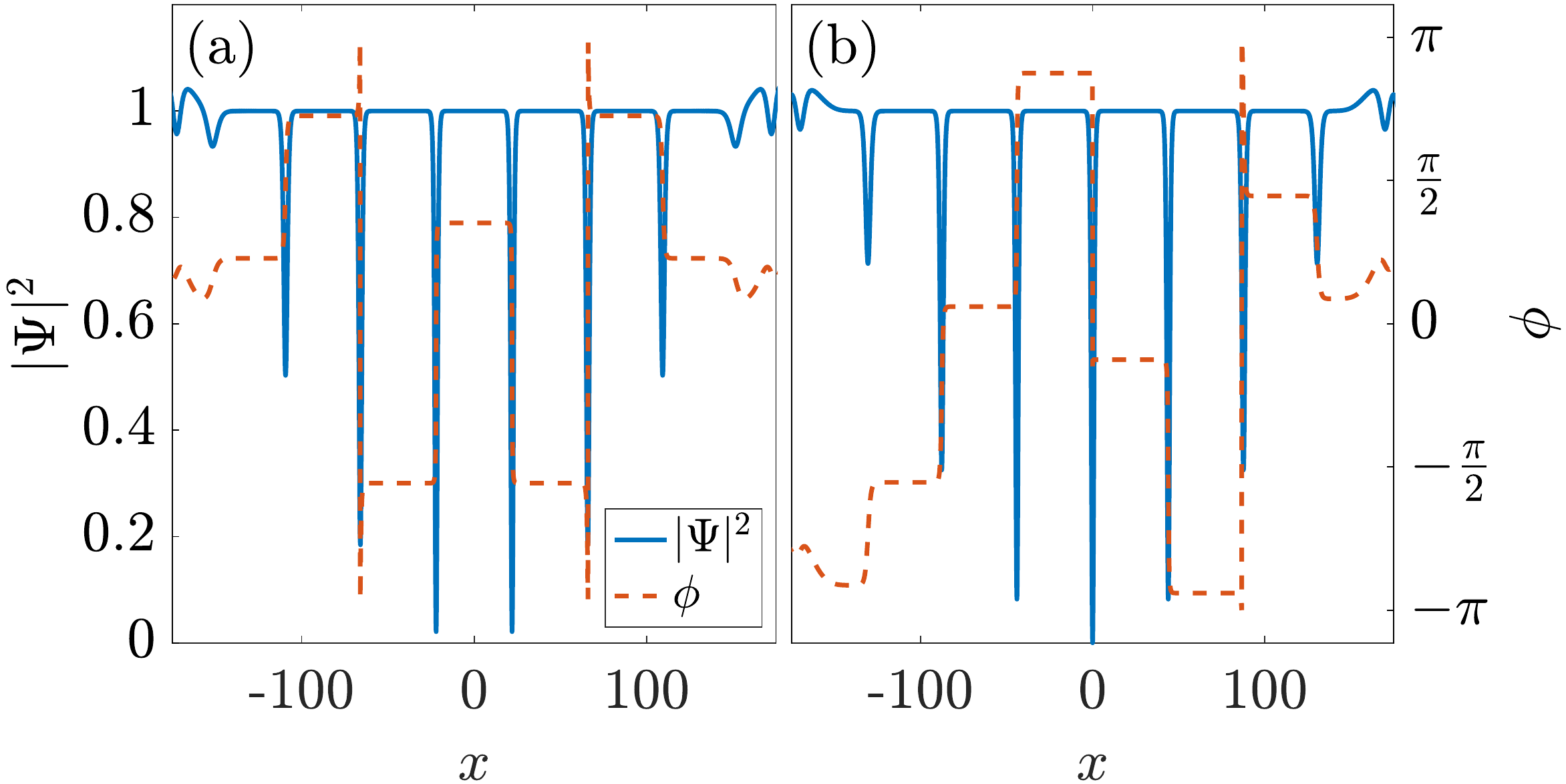}
\caption{Left axes: Profile snapshots of the density, $|\Psi|^2$, at $t=150$ showcasing the generated dark solitons 
for (a) IP-IRP and (b) OP-IRP initial conditions. In both cases $|u|=1$ and $a=5$, resulting in three pairs of dark solitons 
being formed in (a) and three pairs and a central black soliton in (b).
Right axes: Snapshots of the corresponding phase, $\phi$, 
(see legend) for (a) an IP-IRP and (b) an OP-IRP 
illustrating the characteristic phase-jump occurring at each of the dark soliton minima.
Phase-shifts $0 < \Delta\phi < \pi$ correspond to moving (gray) solitons, and the maximum phase-shift $\Delta\phi=\pi$ 
belongs to the black soliton centered at $x=0$ in the OP case (b).}
\label{fig:Ds}
\end{figure}

Figures~\ref{fig:Ds}(a) and \ref{fig:Ds}(b) illustrate profile snapshots (at $t=150$) 
of the density, $|\Psi|^2$, for IP- and OP-IRPs, respectively.
As per our discussion above, an even number of dark solitons is expected and indeed observed for an IP-IRP 
[see Fig.~\ref{fig:Ds}(a)]. In particular, 
for an initial amplitude $|u|=1$ and half-width $a=5$, 
three pairs of dark solitons symmetrically placed around 
the origin ($x=0$) are clearly generated.
On the other hand,
for an OP-IRP an odd number of solitons occurs, consisting of three
pairs of dark states
formed symmetrically around $x=0$ and an isolated black soliton residing at $x=0$ [see Fig.\ref{fig:Ds}(b)].
In both cases, by inspecting the relevant phase, $\phi$, the characteristic phase-jump, $\Delta\phi$, located right at the dark 
density minima, expected for each of the nucleated dark states can be clearly inferred 
[see dashed lines in Figs.~\ref{fig:Ds}(a) and \ref{fig:Ds}(b)]. 
Notice that all the solitons formed for both IP- and OP-IRPs are gray (moving) ones since $0<\Delta\phi<\pi$, 
except for the black one shown in Fig.~\ref{fig:Ds}(b) 
which has a phase shift $\Delta\phi=\pi$.

Up to this point we have briefly reviewed the well-known results regarding the controllable generation of multiple dark solitons in 
homogeneous single-component settings.
Below, we focus on the controllable formation of more complex solitonic entities that appear in multi-component BECs.
In this latter context analytical expressions like the ones provided by Eq.~(\ref{eq:w_n_IP}) and Eq.~(\ref{eq:w_n_OP}) are, 
to the best of our knowledge, currently unavailable in the literature
for the initial waveforms considered herein.
Thus, in the following we resort to a systematic numerical investigation aiming at controlling the emergence of more complex solitonic 
structures consisting of multiple solitons of the DB type.
In particular, we initially focus on the simplest case scenario, i.e. a two-component BEC [see Eq.\eqref{eq:CGPE}].
Next in our systematic progression, we consider a three-component mixture [see Eq.(\ref{eq:3-CGPE})]; finally, we turn our attention 
to the true spinorial BEC system [see Eqs.(\ref{eq:spinor_hamiltonian_a})-(\ref{eq:spinor_hamiltonian_b})].
Additionally, and also in all cases that will be examined herein, 
in order to initialize the dynamics, we use as initial condition for the component(s) that 
during the evolution will host multiple dark solitons the IRP wavefunction given by Eq.~(\ref{eq:square_well}).
Furthermore, for the component(s) that during the evolution will host multiple bright states a Gaussian pulse is used.
The latter ansatz is given by 
\begin{equation}
\Psi(x,0)=\sqrt{A}\exp\left[-\frac{1}{2}\kappa^2(x-X_0)^2\right],
\label{eq:gaussian}
\end{equation}
with $A$, $\kappa$ and $X_0$ denoting respectively the amplitude, the inverse width and the center of the Gaussian pulse.
To minimize the emitted radiation during the counterflow process, in the trapped scenarios the following procedure is used.
The multi-component system is initially relaxed to its ground state configuration. For the relaxation process
we use as an initial guess Thomas-Fermi (TF) profiles for 
all the participating components, i.e. $\Psi (x)=\sqrt{\mu-V_i(x)}$.  
Here, $\mu$ denotes the common chemical potential assumed throughout this work 
for all models under consideration.
It is relevant to mention in passing here that the selection of a common
$\mu$ is a necessity (due to the spin-dependent interaction) in the spinor
system, but not in the Manakov case (where it constitutes a simplification
in order to reduce the large number of parameters in the problem).
Additionally, $i=d,b$ indicates the different traps used 
for the participating species. 
In particular, the component(s) that will host during evolution dark solitons is (are) confined in a double-well potential 
that reads~\cite{Reinhardt1997,Theocharis2010}
\begin{equation}
V_d(x)=V(x)+G\exp\left(-x^2/w^2\right).
\label{eq:V_m}
\end{equation}
In Eq.~(\ref{eq:V_m}), $V(x)$ is the standard harmonic potential, 
while $G$ and $w$ are the amplitude and width of the Gaussian barrier used.
Tuning $G$ and $w$ allows us to control the spatial separation of the two condensates. 
We also note in passing that the choice of Eq.~(\ref{eq:V_m}) is based on the standard way 
to induce the counterflow dynamics in single-component BEC experiments~\cite{Weller2008,hoefer2009matter}.
The remaining component(s) that during evolution will host bright solitons 
are trapped solely in a harmonic potential $V_b(x)=\frac{1}{2}\Omega_b^2x^2$, with $\Omega_b>\Omega$.
The latter choice is made in order to reduce the initial spatial overlap between the components which, in turn, facilitates soliton 
generation during the dynamics.
After the above-discussed relaxation process the system is left to dynamically evolve in the common harmonic potential $V(x)$ 
by switching off the barrier in Eq.~(\ref{eq:V_m}), i.e.
setting $G=0$, and also removing $V_b$ by setting $\Omega_b=0$. 

In all cases under investigation, in order to simulate the counterflow dynamics 
of the relevant mixture a fourth-order Runge-Kutta integrator is employed,
and a second-order finite differences method is used for the spatial derivatives.
The spatial and time discretization are $\diff x=0.1$ and $\diff t=0.001$ respectively.
Moreover, unless stated otherwise, throughout this work we 
fix $|u|=1$, $\Delta\phi=0$ [see Eq.~(\ref{eq:square_well})] 
and $A=1$, $\kappa=1$, $X_0=0$ [see Eq.~(\ref{eq:gaussian})].
The default parameters for the trapped scenarios are
$\mu_j=\mu=1$, (with $j$ denoting the participating components) 
$G=5\mu$, $w^2=5$, $\Omega=0.05$ and $\Omega_b=30\Omega$.
We have checked that slight deviations from these parametric
selections do not significantly affect our qualitative observations
reported below.
Additionally, for the spinor BEC system we also fix 
$g_s=-4.6{\times}10^{-3}$. Notice that the chosen value is exactly the ratio 
$\frac{a_2-a_0}{a_0+2a_2}$ that is (in the range)
typically used in ferromagnetic spinor $F=1$ BEC of $^{87}$Rb 
atoms~\cite{Klausen2001,vanKempen2002}. 
However, we note that the numerical findings to be presented below 
are not altered significantly
even upon considering a spinor $F=1$ BEC of $^{23}$Na atoms.

Finally, focusing on $^{87}$Rb BEC systems, 
our dimensionless parameters can be expressed in dimensional form by 
assuming a transversal trapping frequency $\omega_{\perp}=2\pi \times 175$~Hz.
Then all time scales must be rescaled by $8.1$~s and all length scales
by $100~\rm{\mu m}$.
This yields an axial trapping frequency $\omega_{x} \approx 2\pi \times 1.1$~Hz
which is accessible by current state-of-the-art experiments~\cite{Bersano2018}. 
The corresponding aspect ratio is $\omega_{x}/\omega_{\perp}= 5 \times 10^{-3}$
and as such lies within the range of applicability of the 1D GP theory according to the
criterion $N a^4_{\perp}/a^2 a^2_z \gg 1$~\cite{pitaevskii2016bose}. 
Here $a_{\perp}$, and $a_z$ denote respectively the oscillator
length in the transversal and axial direction, while $a$ is the three-dimensional $s$-wave scattering length.

\section{Numerical Results And Discussion} \label{sec:results}

\subsection{Two-Component BEC} \label{subsec:psuedo-spinor}
In this section we present our findings regarding the controlled generation of arrays
of DB solitons and their robust evolution in 
two-component BECs~\cite{Becker2008,Middelkamp2011,Hamner2011,Yan2011,Hoefer2011,Yan2012}. 
To induce the counterflow dynamics we utilize the methods introduced in Sec.~\ref{sec:model_setup} B. 
Before delving into the associated dynamics we should first recall that in the integrable limit, i.e. $g_n=1$ and $V(x)=0$,
the system of Eqs.~(\ref{eq:CGPE}) admits an exact DB soliton solution. The corresponding DB waveforms 
read~\cite{Yan2011,biondini2016three,katsimiga2017dark,katsimiga2017stability,katsimiga2018dark}
\begin{eqnarray}
\Psi_d (x,t) &=& \Big[\nu \tanh\left[ \mathcal{D} \left (x-x_0(t)\right) \right]+i\lambda \Big]e^{-it},
\label{eq:DB_d} \\
\Psi_b (x,t) &=& \eta \sech\left[\mathcal{D} \left(x-x_0(t)\right) \right] e^{\left[ ikx+i\varphi(t) \right]},
\label{eq:DB_b}
\end{eqnarray}
and are subject to the boundary conditions $|\Psi_d|^2\rightarrow 1$ and 
$|\Psi_b|^2\rightarrow 0$ as $|x|\rightarrow \infty$, in the dimensionless units adopted herein.  
In Eqs.~(\ref{eq:DB_d})-(\ref{eq:DB_b}) $\Psi_d$ ($\Psi_b$) is the wavefunction of the dark (bright) soliton component. 
In the aforementioned solutions, $\nu$ and $\eta$ are the amplitudes of the dark and the bright soliton respectively, 
while $\lambda$ sets the velocity of the dark soliton.
Furthermore, $\mathcal{D}$ denotes the common --across components--
  inverse width
parameter and $x_0(t)$, which will be traced numerically later on, 
refers to the center position of the DB soliton (see also our discussion below). 
Additionally, in the above expressions $k=\mathcal{D}\left(\lambda/\nu\right)$ 
is the constant wavenumber of the bright soliton associated with the DB soliton's velocity, 
and $\varphi(t)$ is its phase.
Inserting the solutions of Eqs.~(\ref{eq:DB_d})-(\ref{eq:DB_b}) in the system of 
Eqs.~(\ref{eq:CGPE}) leads to the following conditions that the DB soliton parameters must satisfy for the above solution
to exist 
\begin{eqnarray}
\mathcal{D}^2 &=& \nu^2-\eta^2 ,
\label{eq:DB_D}
\\
\dot{x}_0 &=& \mathcal{D}\frac{\lambda}{\nu},
\label{eq:DB_v}
\end{eqnarray}
where $\dot{x}_0$ is the DB soliton velocity.
Through the normalization of $\Psi_b$ we can connect the number of particles of the bright component, $N_b$, with $\eta$ and $\mathcal{D}$
\begin{equation}
N_b=\int |\Psi_b(x,t)|^2 \text{d}x=\frac{2\eta^2}{\mathcal{D}}.
\label{eq:DB_N}
\end{equation}
In the following we will use the aforementioned conditions, namely Eqs.~(\ref{eq:DB_D})-(\ref{eq:DB_v}),
not only to verify the nature of the emergent states but also 
to compare the trajectories of the evolved DBs to the analytical prediction provided by Eq.~(\ref{eq:DB_v}).
Moreover, by making use of Eq.~(\ref{eq:DB_N}) we will further estimate 
the number of particles hosted in the bright soliton component of the mixture.

\begin{figure}[t]
\centering
\includegraphics[width=0.48\textwidth]{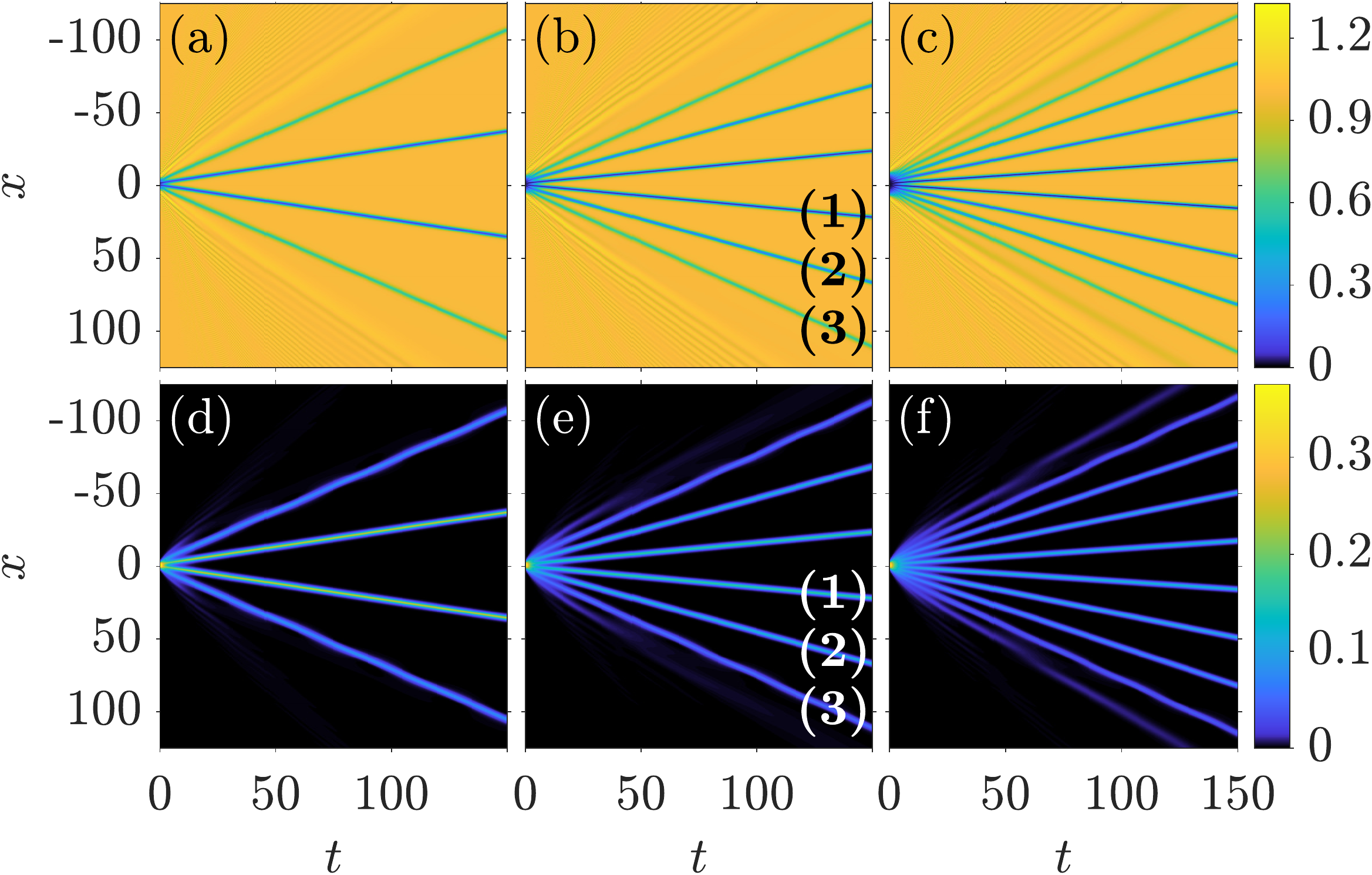}
\caption{Spatio-temporal evolution of the density $\abs{\Psi_1}^2$ ($\abs{\Psi_2}^2$) 
of the first (second) component upon varying the 
half-width $a$ of the initial IRP.
From left to right $a=3$, $a=5$ and $a=7$, allowing the generation of four [(a)-(d)], 
six [(b)-(e)], and ten [(c)-(f)] DB solitons, respectively.
In all cases, top (bottom) panels illustrate the formation of dark (bright) solitons in the first (second) 
component of the two-component system.
Labels $(1)$-$(3)$ introduced in panels $(b)$, $(e)$ number the DB solitons discussed in Table \ref{tab:DB_char}.}
\label{fig:DB}
\end{figure}

The outcome of the counterflow process for different variations of the half-width $a$ of the initial IP-IRP 
is illustrated in Figs.~\ref{fig:DB}(a)-\ref{fig:DB}(f).   
In particular, in all cases depicted in this figure, the spatio-temporal evolution of the densities, $|\Psi_j|^2$ (with $j=1,2$), 
of both components for propagation times up to $t=150$ is presented.
It is found that from the very early stages of the dynamics the interference fringes in the first component evolve into 
several dark soliton states being generated in this component. E.g. four dark solitons can be readily seen 
in Fig.~\ref{fig:DB}(a) for $a=3$. 
The nucleation of these dark states leads in turn to the emergence, 
via the confinement of the spreading Gaussian pulse,
also of four bright solitons in the second component of the binary mixture [Fig.~\ref{fig:DB}(d)].
The latter bright waveforms are created in each of the corresponding dark minima, and are subsequently waveguided
by their dark counterparts.   
The robust propagation of the newly formed array of DB solitons is illustrated for times up to $t=150$.
Importantly here, we were able to showcase that by tuning the half-width of the initial IRP a controllable formation 
of arrays of DB solitons can be achieved. 
In particular, it is found that upon increasing the initial half-width of the IP-IRP leads to a larger number of 
DB solitons being generated. 
Indeed, as shown in Figs.~\ref{fig:DB}(b) and \ref{fig:DB}(e) six  
DB states are formed for $a=5$, while for $a=7$ the resulting array consists of 
ten DB solitons as illustrated in Figs.~\ref{fig:DB}(c) and \ref{fig:DB}(f). 
We should remark also here that since an IP-IRP is utilized only an even number of DB solitons 
is expected and indeed observed in all of the aforementioned cases.
This result is in line with the analytical predictions discussed in the single-component scenario [see also Eq.~(\ref{eq:w_n_IP})].
 
Moreover, to verify that indeed the entities formed are DB solitons we proceed as follows.
Firstly, upon fitting it is confirmed that the evolved dark and bight states have the standard
$\tanh$- and $\rm{sech}$-shaped waveform respectively [see Eqs.~(\ref{eq:DB_d})-(\ref{eq:DB_b})].  
Then, by monitoring during evolution a selected DB pair we measure the amplitudes 
$\nu$ and $\eta$ of the dark and the bright constituents, respectively. 
Having at hand the numerically obtained amplitudes we then use the analytical expressions stemming from the single DB 
soliton solution, namely Eqs.~(\ref{eq:DB_D})-(\ref{eq:DB_N}). 
In this way estimates of the corresponding DB trajectory
as well as the number of particles, $N_b$, hosted in the selected bright soliton are extracted.
Via the aforementioned procedure and e.g. for the closest to the origin ($x=0$) right moving DB solitary wave
labeled as (1) and shown in Figs.~\ref{fig:DB}(b)-\ref{fig:DB}(e) 
it is found that $N_b=0.3611$ while the numerically obtained value is $N^{num}_b=0.3607$.
Notice that the deviation between the semi-analytical calculation and the numerical one is less than $1\%$.
To have access to $N^{num}_b$  we simply integrated $|\Psi_2|^2$ within a small region around the center of the bright part 
of the selected DB pair. Additionally, for the same DB pair $\dot{x}_0=0.1467$ while $\dot{x}^{num}_0=0.1495$.

After confirming that all entities illustrated in Figs.~\ref{fig:DB}(a)-\ref{fig:DB}(f)
are indeed DB solitons, with each of the resulting DBs following the analytical predictions of 
Eqs.~(\ref{eq:DB_D})-(\ref{eq:DB_N}), we next consider different parametric variations.
In particular, we will investigate modifications in the DB soliton
characteristics when the number of the nucleated DB states is held fixed.
To this end, below we fix $a=5$ and we then vary within the interval $[0.5, 2]$ one of the
following parameters at a time: $|u|$, $A$, $\kappa$. 

Before proceeding, two important remarks are of relevance at this point.
(i) Fixing $a=5$ is not by itself sufficient to {\it a priori} ensure that a fixed number of
DB solitons will be generated via the interference process. 
This is due to the fact that the number of solitons formed is proportional to $a$ and 
$|u|$ as detected by Eq.~(\ref{eq:w_n_IP}).
This is the reason for restricting ourselves to the aforementioned interval in terms of $|u|$ ($|u| \in [0.5, 2]$).
This selection
leads to the formation of an array consisting of only six DB solitons
like the ones shown in Figs.~\ref{fig:DB}(b) and \ref{fig:DB}(e) for $A=\kappa=1$.
(ii) Additionally here, variations of either $A$, or $\kappa$ could in principle affect the bright soliton 
formation, however we have not found this to be the case in our intervals
of consideration.

Taking advantage of the symmetric formation of these six DB structures 
in the analysis that follows we will
focus our attention to the three, i.e. (1), (2), and (3), right moving with respect to $x=0$ 
DB solitons shown in Figs.~\ref{fig:DB}(b) and \ref{fig:DB}(e).
\begin{table}[t]
  \def\arraystretch{1.5}
  \centering  
  \begin{tabular}{c | ccccc | ccccc | ccccc }  
  \toprule
  [0.5, 2]
       & \multicolumn{5}{c}{$\abs{u}  \uparrow$}
       & \multicolumn{5}{ | c | }{$A  \uparrow$}
       & \multicolumn{5}{c}{$\kappa \uparrow$} \\
   \toprule
   DB
       & $\nu$		& $\eta$	& $D$		& $n_b$ 	& $\dot{x}_0$
       & $\nu$		& $\eta$	& $D$		& $n_b$ 	& $\dot{x}_0$
       & $\nu$		& $\eta$	& $D$		& $n_b$ 	& $\dot{x}_0$ \\ 
   \hline
   (1) & $\uparrow$	& $\uparrow$	& $\uparrow$	& $\uparrow$	& $\uparrow$
       & $\downarrow$	& $\uparrow$	& $\downarrow$	& $\downarrow$	& $\downarrow$
       & $\uparrow$	& $\downarrow$	& $\uparrow$	& $\downarrow$	& $\downarrow$ \\
   (2) & $\uparrow$	& $\uparrow$	& $\uparrow$	& $\uparrow$	& $\uparrow$
       & $\downarrow$	& $\uparrow$	& $\downarrow$	& $\downarrow$	& $\downarrow$
       & $\uparrow$	& $\downarrow$	& $\uparrow$	& $\downarrow$	& $\downarrow$ \\
   (3) & $\uparrow$	& $\uparrow$	& $\uparrow$	& $\downarrow$	& $\uparrow$
       & $\downarrow$	& $\uparrow$	& $\downarrow$	& $\uparrow$	& $\downarrow$
       & $\uparrow$	& $\downarrow$	& $\uparrow$	& $\uparrow$	& $\downarrow$ \\  
   \toprule
  \end{tabular}   
  \caption{Changes in the DB soliton characteristics upon considering different variations of the systems' parameters 
  for fixed $a=5$.
  Here, $(1)$ [$(3)$] refers to the most inner [outer] DB solitons [see Figs.~\ref{fig:DB}(b) and \ref{fig:DB}(e)]. 
  The top row indicates the distinct variations, namely of each 
  $|u|$, $A$, and $\kappa$, performed separately within the interval $[0.5, 2]$. 
  The second row contains the soliton characteristics such as the dark, $\nu$, and bright, $\eta$, amplitudes. 
  Also shown are the inverse width, $\mathcal{D}$, the normalized number of particles, $n_b$, hosted in each 
  of the bright solitons formed in the second component 
  of the mixture, and the velocity, $\dot{x}_0$, of the DB pair. 
  $\uparrow$ ($\downarrow$) arrows  indicate an increase (decrease) of the corresponding quantity.}
  \label{tab:DB_char}
\end{table}  
The effect that different parametric variations have on the characteristics 
of these three DB solitons are summarized in Table~\ref{tab:DB_char}. 
In particular in this table, the arrow $\uparrow$ ($\downarrow$) indicates an 
increase (decrease) of the corresponding soliton characteristic as one of the parameters 
$|u|$, $A$ and $\kappa$ is increased within the chosen interval.
In general, it is found that as the amplitude, $|u|$, of the initial IP-IRP 
increases the amplitudes, $\nu$, $\eta$, of all three DB structures increase
as well [see the second column in Table~\ref{tab:DB_char}].   
Also, the resulting DB states are found to be narrower 
(larger inverse width $\mathcal{D}$) and faster (larger $\dot{x}_0$).
However, the normalized number of particles, $n_b$,
hosted in each of the bright soliton constituents is found to increase for the two 
innermost DB states [i.e. (1) and (2)] while it decreases for the outer one [i.e. (3)].
For instance, for the inner DB wave labeled (1) shown in the first column of Table~\ref{tab:DB_char},
$n_b$ is found to be $n_b = 0.196$ for $|u|= 0.5$, while $n_b = 0.204$ for $|u|= 1$.
Thus, the symbol $\uparrow$ is used to describe the increasing tendency of $n_b$ 
[see the second column of Table~\ref{tab:DB_char}]. 
We defined $n_b$ according to $n_b=N^{num}_b/N_2$ with $N_2=\int|\Psi_{2}|^2\text{d}x$ being the total number of 
particles in the second component of the binary mixture.
For comparison here, for the outer DB soliton labeled (3)
$n_b = 0.092$ for $|u|= 0.5$ while $n_b = 0.075$ for $|u|= 1$ and thus a
symbol $\downarrow$ 
is introduced [see again the second column in Table~\ref{tab:DB_char}].

On the contrary upon increasing the amplitude, $A$, of the initial Gaussian pulse [see Eq.~(\ref{eq:gaussian})]
the amplitudes of all dark (bright) solitons for all three DB pairs decrease (increase), thus a decrease of the corresponding 
inverse width results in wider and slower soliton pairs [see the third column in Table~\ref{tab:DB_char}].
Moreover, $n_b$ is found to decrease for the two inner DB pairs while it increases for the outer one.
Variations of the inverse width, $\kappa$, of the Gaussian pulse have more or less the opposite to the above-described effect.
As $\kappa$ increases, the resulting dark (bright) states have larger (smaller) 
amplitudes for all three DB pairs but the solitons are narrower and slower 
[see the fourth column in Table~\ref{tab:DB_char}]. 
Recall that narrower does not directly imply faster states since 
the amplitude of the generated dark solitons is also involved [see Eqs.~(\ref{eq:DB_D}), (\ref{eq:DB_v})].
Also in this case $n_b$ increases for the outer DB pair [see the fourth column in Table~\ref{tab:DB_char}].
Finally, we also considered different displacements, $X_0$, of the initial Gaussian pulse 
within the interval $[0, 7.5]$. 
A behavior similar to the aforementioned $\kappa$ variation
is observed. However, the produced solitons are found to be asymmetric for $X_0 \neq 0$
due to the asymmetric positioning of the two components. 
On the other hand, for $X_0 \geq a$ ($a=5$) we never observe DB soliton generation.

Having discussed in detail the homogeneous system,
we next turn our attention to the harmonically confined one [see Eq.~(\ref{eq:CGPE})]. 
Recall that in this case the initial guesses used for both components of the binary mixture are TF profiles. 
The first component is initially confined in the double-well potential $V_d(x)$  
with the width $w$ of the barrier controlling the spatial separation of the two parts of the condensate [see 
Eq.~(\ref{eq:V_m})].
The corresponding second component  
is in turn trapped in the harmonic potential $V_b(x)$ (see Sec.~\ref{sec:model_setup} B). 
After relaxation the two-component system is left to dynamically evolve in the common parabolic trap $V(x)$. 

\begin{figure}[t]
\centering
\includegraphics[width=0.48\textwidth]{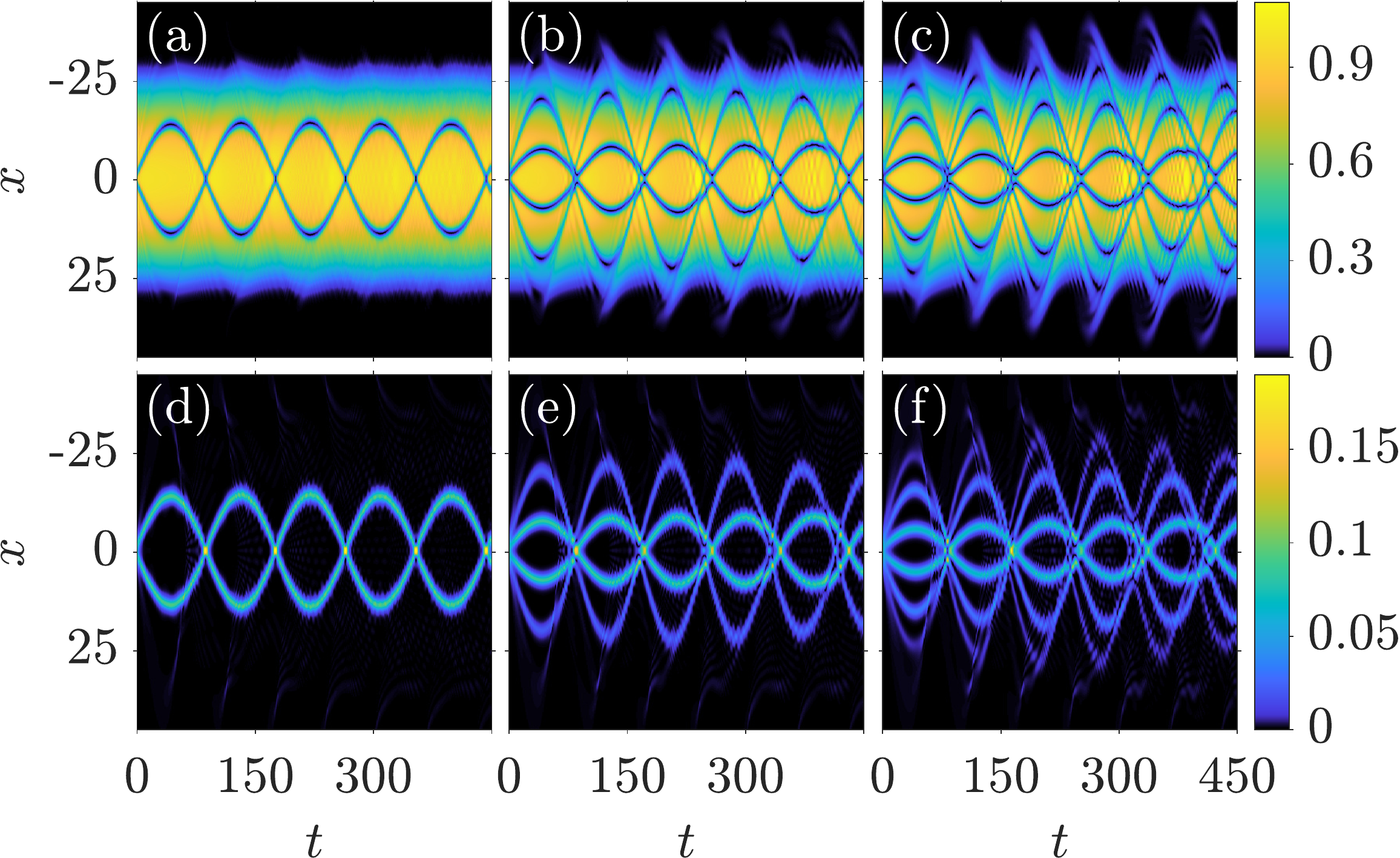}
\caption{Spatio-temporal evolution of the density $\abs{\Psi_1}^2$ ($\abs{\Psi_2}^2$) of the first (second) component 
in the trapped scenario upon varying the width of the double-well barrier $w$ used for the preparation of the initial state.
From left to right $w^2=1$, $w^2=5$ and $w^2=10$, 
allowing the generation of two [(a)-(d)], four [(b)-(e)], and six [(c)-(f)] DB solitons respectively.
In all cases, top (bottom) panels illustrate the formation of dark (bright) solitons in the first (second) 
component of the two-component system.}
\label{fig:DB_trap}
\end{figure} 

In line with our findings for the homogeneous setting, also here a desirable number of DB solitons 
can be achieved by properly adjusting either $w$ or the chemical potentials $\mu_i$ (with $i=1,2$) of the binary mixture.
Note that in this latter case, the amplitude of the system is directly related to $\mu$ [see Eq.~(\ref{eq:w_n_IP})].
In both cases, it is found that an increase of $w$ or $\mu$ results to more DB solitons being generated. 
In particular, Figs.~\ref{fig:DB_trap}(a)-\ref{fig:DB_trap}(c) [Figs.~\ref{fig:DB_trap}(d)-\ref{fig:DB_trap}(f)] 
illustrate the dynamical evolution of the density, $|\Psi_1|^2$ ($|\Psi_2|^2$), of the first (second) component 
of the mixture upon increasing $w$. An array consisting of two, four and six DB solitons pairs can be observed for $w^2=1$, 
$w^2=5$ and $w^2=10$ respectively. 
In all cases depicted in this figure the DB states are formed from the very early stages of the 
dynamics. After their formation the states begin to oscillate within the parabolic trap. 
Monitoring their propagation for evolution times up to $t=450$, it is found that while coherent oscillations are 
observed for the two DB case [see Figs.~\ref{fig:DB_trap}(a), \ref{fig:DB_trap}(d)], this picture is altered for larger DB 
soliton arrays.
In the former case measurements of the oscillation frequency, $\omega_{osc}$, 
verify that it closely follows the analytical 
predictions for the single DB soliton. 
Namely, $\omega_{osc}=\Omega^2\left(\frac{1}{2}-\frac{\chi}{\chi_0}\right)$, with 
$\chi=N_2/\sqrt{\mu}$ and $\chi_0=8\sqrt{1+\left(\frac{\chi}{4}\right)^2}$~\cite{Busch2001,Yan2011}.
For instance, our semi-analytical calculation stemming from the aforementioned theoretical prediction 
gives $\omega_{osc}=34.3 \times 10^{-3}$, while direct measurements from our numerical 
simulations provide $\omega_{osc}^{num}=35.3 \times 10^{-3}$. 
This represents a $3\%$ discrepancy, which can be attributed to the interaction 
of the solitons both with one another but also
with the background 
excitations, with the latter having the form of sound waves.
Additionally, it should be noted that the theoretical prediction is
valid in the large $\mu$ limit (which may be partially responsible
for the relevant discrepancy). 
However, for larger DB soliton arrays the number of collisions is higher and the background density is more excited,
as can be deduced by comparing Figs.~\ref{fig:DB_trap}(a), \ref{fig:DB_trap}(d)  
to Figs.~\ref{fig:DB_trap}(b), \ref{fig:DB_trap}(e) and Figs.~\ref{fig:DB_trap}(c), \ref{fig:DB_trap}(f).
Importantly here the generated DB states are of different mass and thus each DB soliton oscillates 
with its own $\omega_{osc}$. It is this mass difference that results in 
the progressive ``dephasing'' observed during evolution. 
Notice also that in all cases illustrated in the aforementioned figures 
the outer (faster) DB solitons are the ones that are affected the most.
The above effect is enhanced 
for larger initial separations $w$ [compare Figs.~\ref{fig:DB_trap}(b), \ref{fig:DB_trap}(e)  
to Figs.~\ref{fig:DB_trap}(c), \ref{fig:DB_trap}(f)],
leading to discrepancies up to $11.6\%$ between $\omega_{osc}$ and $\omega^{num}_{osc}$ observed for the
outermost DB pair shown in Figs.~\ref{fig:DB_trap}(c), \ref{fig:DB_trap}(f).
\begin{figure}[t]
\centering
\includegraphics[width=0.48\textwidth]{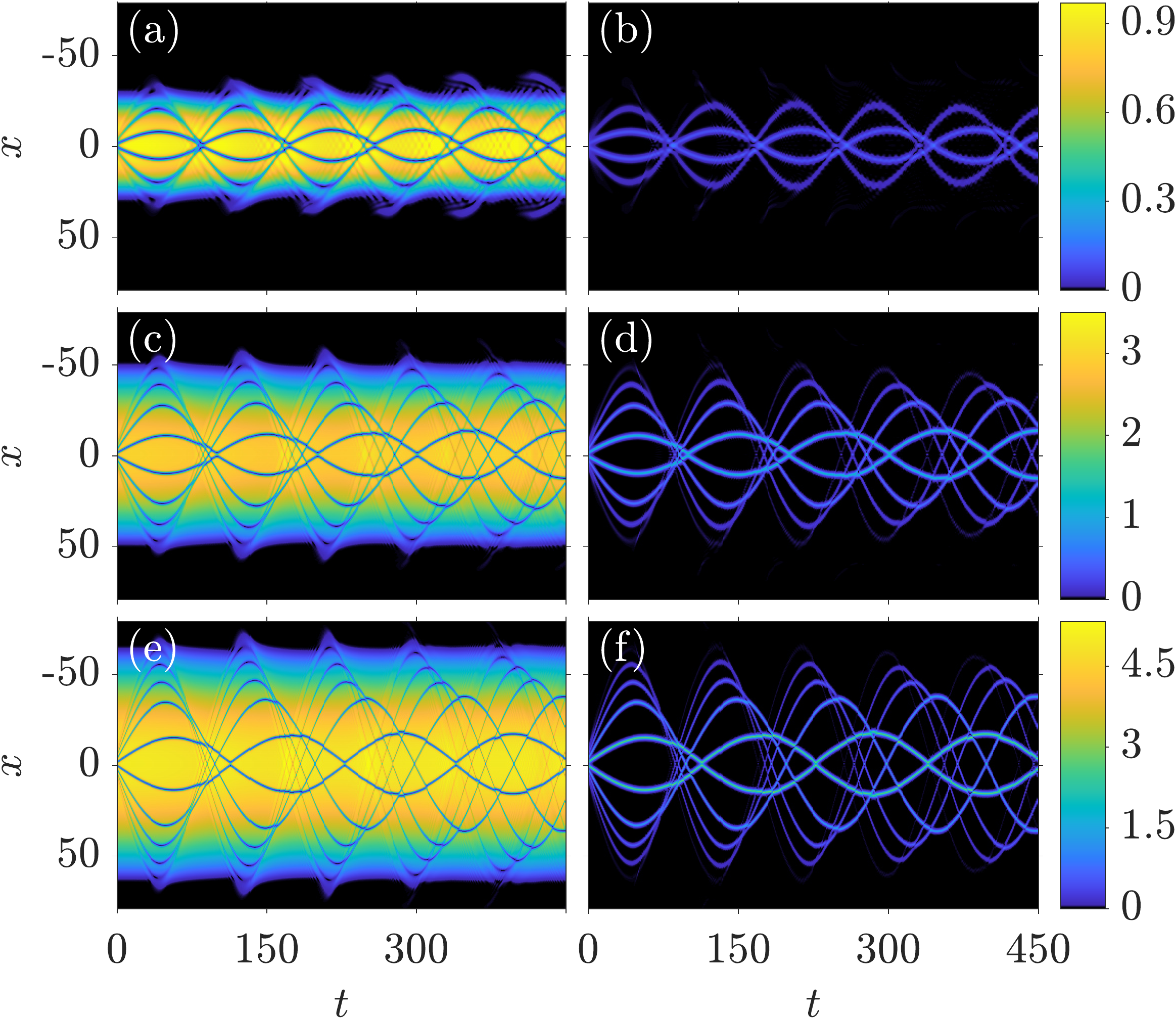}
\caption{Spatio-temporal evolution of the density $\abs{\Psi_1}^2$ ($\abs{\Psi_2}^2$) 
of the first (second) component in the trapped scenario 
upon varying the chemical potential $\mu$ while fixing $w^2=5$.
From top to bottom $\mu=1$, $\mu=3$ and $\mu=5$, 
leading to the emergence of four [(a)-(b)], six [(c)-(d)], and eight [(e)-(f)] DB solitons 
respectively.
In all cases, left (right) panels illustrate the formation of dark (bright) solitons 
in the first (second) component of the two-component system.}
\label{fig:DB_trap_mu} 
\end{figure}

As mentioned above, besides $w$ also the chemical potential $\mu$ serves as a controlling parameter. 
Indeed, by inspecting the spatio-temporal evolution of the densities, $|\Psi_j|^2$ (with $j=1,2$), shown in 
Figs.~\ref{fig:DB_trap_mu}(a)-\ref{fig:DB_trap_mu}(f) for fixed $w^2=5$ it becomes apparent that  
increasing $\mu$ leads to an increased number of DB solitons
being generated. 
Four, six and eight DB solitons
are seen to be nucleated for  
$\mu=1$, $\mu=3$ and $\mu=5$ respectively, and to propagate 
within the BEC medium for long evolution times.
Notice that Figs.~\ref{fig:DB_trap_mu}(a), \ref{fig:DB_trap_mu}(b) 
are the same as Figs.~\ref{fig:DB_trap}(b), \ref{fig:DB_trap}(e).
Increasing the system size reduces the impact that the radiation expelled
(when matter-wave interference takes place) has on the resulting DB states, as can be deduced 
by comparing Figs.~\ref{fig:DB_trap_mu}(c), \ref{fig:DB_trap_mu}(d) to Figs.~\ref{fig:DB_trap}(c), \ref{fig:DB_trap}(f).
Indeed, further measurements of $\omega_{osc}$ reveal that the maximum 
discrepancy observed for the outermost DB solitons when $\mu=1$ 
[see Figs.~\ref{fig:DB_trap_mu}(a),\ref{fig:DB_trap_mu}(b)] is of about $8.5\%$, 
while upon increasing $\mu$ the discrepancy is significantly reduced.
The latter reduction is  attributed to the fact that for larger $\mu$
the asymptotic prediction of $\omega_{osc}$ is progressively more accurate.
More specifically, for $\mu=5$ we obtain a discrepancy of 
only $0.3\%$ for the third, with respect to $x=0$, DB soliton pair 
shown in Figs.~\ref{fig:DB_trap_mu}(e), \ref{fig:DB_trap_mu}(f).
Yet still, the emergent DB states have different periods of oscillation,
leading in turn to several collision events taking place during 
evolution. Nevertheless, in all cases presented above a common feature
of the solitary waves is that they survive throughout our computational
horizon.

\begin{figure}[t]
\centering
\includegraphics[width=0.48\textwidth]{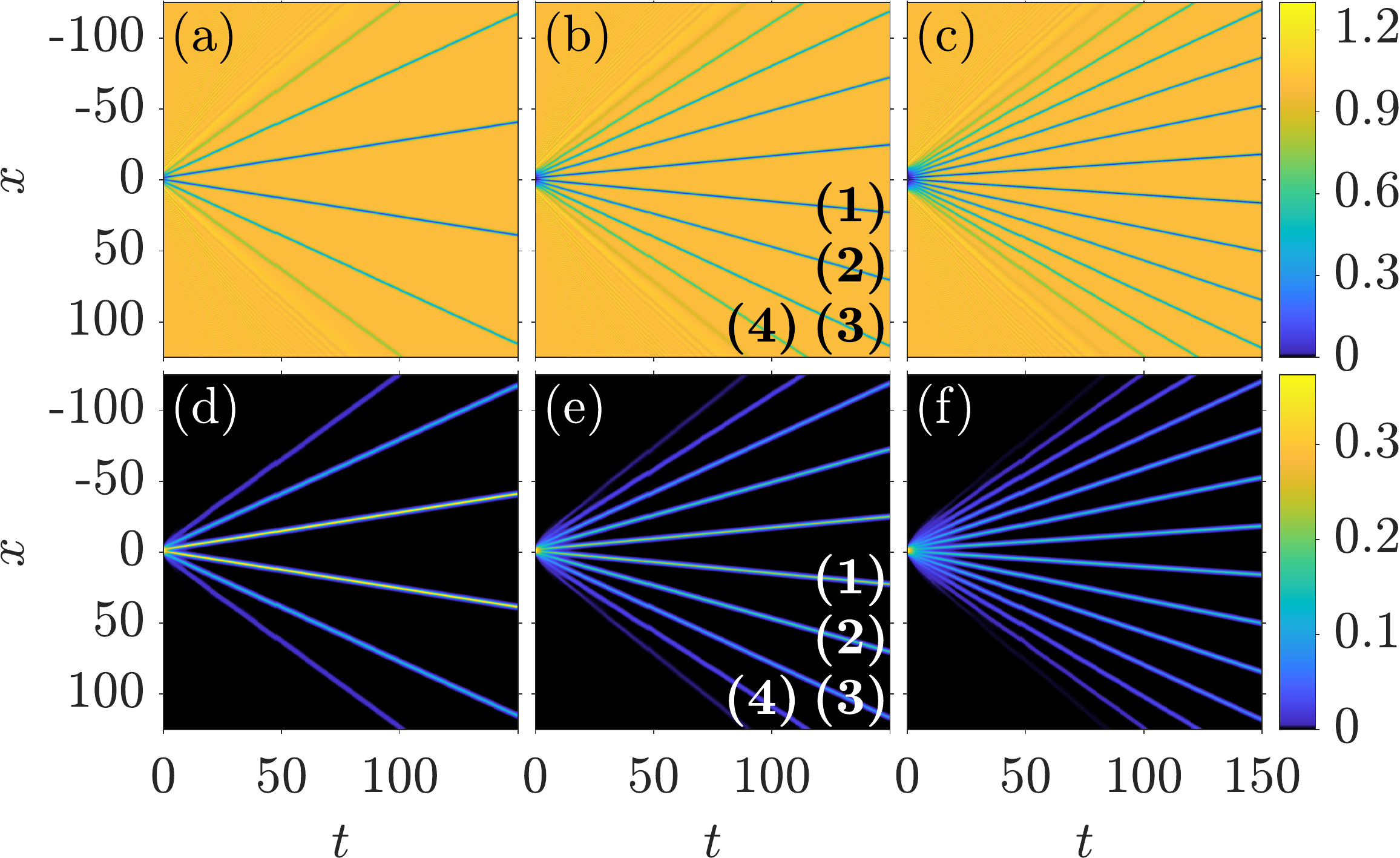}
\caption{Spatio-temporal evolution of the density $|\Psi_{+1}|^2$ ($|\Psi_{-1}|^2$) of the $m_F={+1}$ ($m_F={-1}$) 
component upon varying the half-width $a$ of the initial IRP.
From left to right $a_{+1}=a_{0}=3$, $a_{+1}=a_{0}=5$ and $a_{+1}=a_{0}=7$, resulting in the nucleation 
of six [(a), (d)], eight [(b), (e)], and twelve [(c), (f)] DDB solitons respectively.
In all cases, top (bottom) panels illustrate the formation of dark (bright) solitons in the $m_F={+1}$ ($m_F={-1}$) component 
of the three-component system. 
Since the evolution of the $m_F={0}$ component is the same as the one depicted for the $m_F={+1}$, 
only the two components that differ from one another are illustrated.
The labels, (1)-(4) introduced in [(b), (e)] number the DDB solitons that are discussed in Table~\ref{tab:DDB_char}. }
\label{fig:DDB} 
\end{figure}

\subsection{Three-component BEC mixtures} \label{subsec:three-component}
Now,  we increase the complexity of the system  
by adding yet another component to the previously discussed two-component mixture. 
Namely we consider a three-component mixture consisting of three different hyperfine states
of the same alkali isotope such as $^{87}$Rb.
We aim at revealing the DB soliton complexes
that arise in such a system and their controllable formation via the interference processes introduced in 
Sec.~\ref{subsec:setup_homo}.  
From a theoretical point of view, such a three-component BEC mixture is described by a system of three coupled GPEs
(see Eqs.~(\ref{eq:3-CGPE}) in Sec.~\ref{subsec:models}), i.e., one for each of the participating $m_F=+1,0,- 1$ components. 

To begin our analysis, we start with the integrable version 
of the problem at hand. 
Namely, we fix $g_n=1$ and we set $V(x)=0$ in the corresponding Eqs.~(\ref{eq:3-CGPE}).
This homogeneous mixture admits exact solutions in the form of DDB and DBB solitons 
as it was rigorously proven via the inverse scattering method~\cite{biondini2016three}.
In the following, we will attempt to produce in a controlled fashion arrays consisting of these types of
soliton compounds.
We further note that in the numerical findings to be presented below
the abbreviations in the form XYZ (with X,Y,Z=D or B) reflect the $m_F=+1,0,-1$ order. 
E.g. a DDB abbreviation indicates that dark solitons are generated in the $m_F=+1,0$ components 
while bright solitons are generated in the $m_F=-1$ component of the mixture.

As it was done in the two-component setting, in order to generate a DDB configuration
the counterflow dynamics is performed by two of the participating hyperfine components.
Recall that dark solitons in each hyperfine state emerge via the destructive interference 
that takes place at the origin where the two spatially separated sides of the initial IP-IRP collide.   
Specifically, the initial ansatz used for the
$m_F=+1,0$ states is provided by Eq.~(\ref{eq:square_well}) and the corresponding ansatz for the $m_F=-1$ component
is the Gaussian of Eq.~(\ref{eq:gaussian}). 
It turns out that we can again tailor the number of nucleated DDB solitons 
by manipulating the half-width, $a_{m_F}$ (with ${m_F}=+1, 0$), of the initial IP-IRP.
To showcase the latter in Figs.~\ref{fig:DDB}(a)-\ref{fig:DDB}(f)
we present the outcome of the distinct variations of $a_{m_F}$.
Notice that as $a_{m_F}$ increases arrays consisting of a progressively 
larger number of DDB solitons are formed. 
Namely, $a_{+1}=a_{0}=3$ results in an array of six DDB solitons~[Figs.~\ref{fig:DDB}(a), \ref{fig:DDB}(d)].
Accordingly, when $a_{+1}=a_{0}=5$ the nucleation of eight DDB waveforms is observed 
[see Figs.~\ref{fig:DDB}(b), \ref{fig:DDB}(e)],
while twelve such states occur for $a_{+1}=a_{0}=7$ [Figs.~\ref{fig:DDB}(c), \ref{fig:DDB}(f)].
In all of the above cases the spatio-temporal evolution of the densities 
$|\Psi_{+1}|^2$, and $|\Psi_{-1}|^2$, are shown in the top and bottom panels of Fig.~\ref{fig:DDB} respectively.
The resulting propagation of the ensuing DDB states is monitored for evolution times up to $t=150$.
Moreover, only the $m_F=\pm 1$ components are depicted in the aforementioned figure. 
This is due to the fact that the evolution of the $m_F=0$ component is
essentially identical to the one shown for the $m_F=+1$ component.
\begin{figure}[t]
\centering
\includegraphics[width=0.48\textwidth]{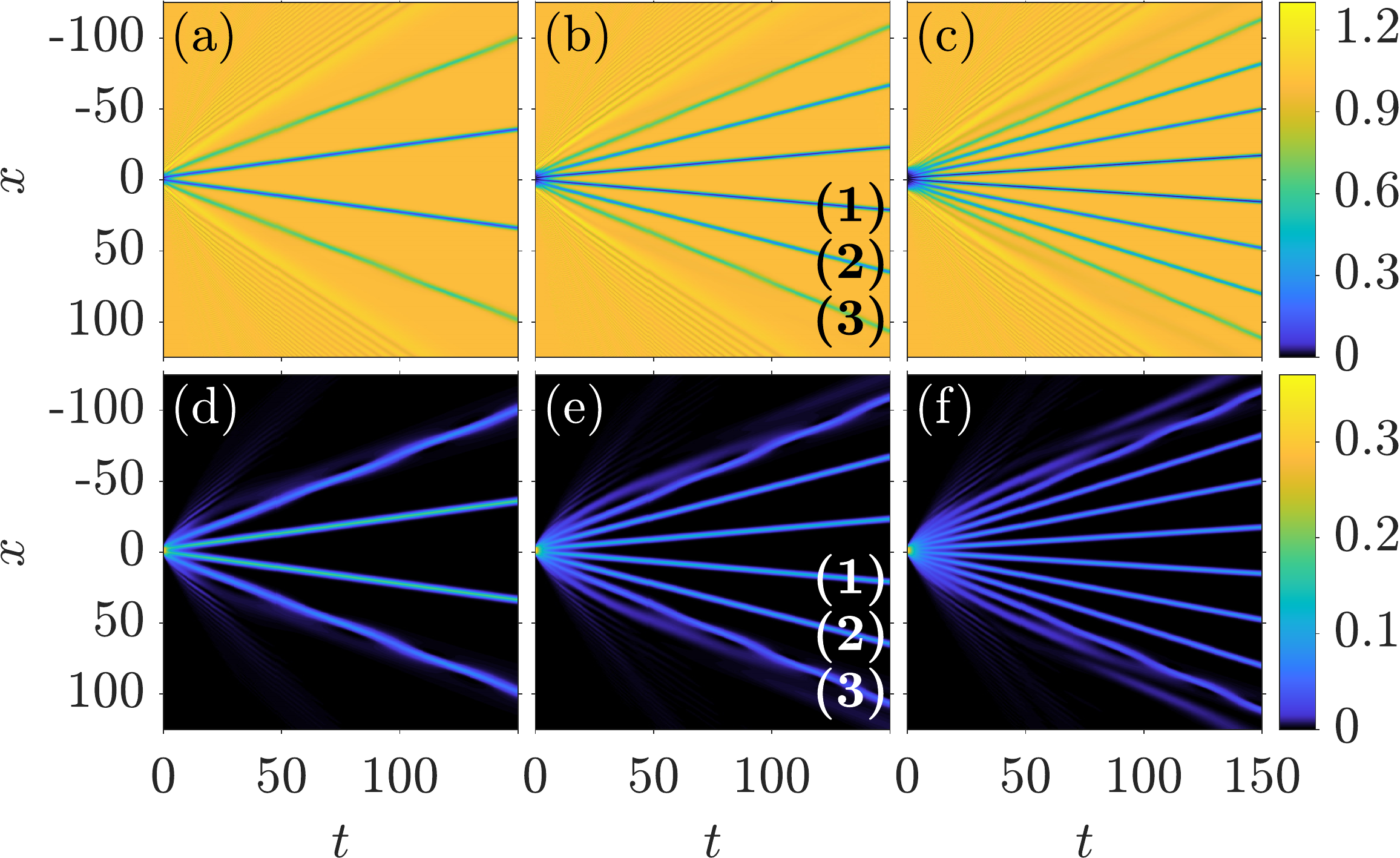}
\caption{Same as in Fig.~\ref{fig:DDB} but showcasing the generation of DBB solitons.
In this case, from left to right $a_{+1}=3$, $a_{+1}=5$ and $a_{+1}=7$, 
allowing the generation of four [(a)-(d)], six [(b)-(e)], 
and eight [(c)-(f)] DBB solitons respectively. 
The labels, (1)-(3) introduced in [(b), (e)] number the DBB solitons that are discussed in Table~\ref{tab:DBB_char}.}
\label{fig:DBB} 
\end{figure}
\begin{table*}
  \def\arraystretch{1.5}
  \centering  
  \begin{tabular}{c | cccccc | cccccc | cccccc} 
  \toprule
  [0.5 , 2]
       & \multicolumn{6}{c}{$\abs{u_0}  \uparrow$}
       & \multicolumn{6}{ | c | }{$A_{-1}  \uparrow$}
       & \multicolumn{6}{c}{$\kappa_{-1} \uparrow$} \\
   \toprule
   DDB
       & $\nu_{+1}$	& $\nu_0$	& $\eta_{-1}$	& $\mathcal{D}$	& $n_b$ 	& $\dot{x}_0$
       & $\nu_{+1}$	& $\nu_0$	& $\eta_{-1}$	& $\mathcal{D}$	& $n_b$ 	& $\dot{x}_0$
       & $\nu_{+1}$	& $\nu_0$	& $\eta_{-1}$	& $\mathcal{D}$	& $n_b$ 	& $\dot{x}_0$ \\ 
   \hline
   (1) & $\uparrow$	& $\uparrow$	& $\uparrow$	& $\uparrow$	& $\uparrow$	& $\uparrow$
       & $\downarrow$	& $\downarrow$	& $\uparrow$	& $\downarrow$	& $\downarrow$	& $\updownarrow$
       & $\uparrow$	& $\uparrow$	& $\downarrow$	& $\uparrow$	& $\downarrow$	& $\downarrow$ \\
   (2) & $\uparrow$	& $\uparrow$	& $\uparrow$	& $\uparrow$	& $\uparrow$	& $\uparrow$
       & $\downarrow$	& $\downarrow$	& $\uparrow$	& $\downarrow$	& $\downarrow$	& $\updownarrow$
       & $\uparrow$	& $\uparrow$	& $\downarrow$	& $\uparrow$	& $\downarrow$	& $\downarrow$ \\
   (3) & $\uparrow$	& $\uparrow$	& $\uparrow$	& $\uparrow$	& $\downarrow$	& $\uparrow$
       & $\downarrow$	& $\downarrow$	& $\uparrow$	& $\downarrow$	& $\uparrow$	& $\updownarrow$
       & $\uparrow$	& $\uparrow$	& $\downarrow$	& $\uparrow$	& $\uparrow$	& $\downarrow$ \\
   (4) & $\uparrow$	& $\uparrow$	& $\uparrow$	& $\uparrow$	& $\downarrow$	& $\uparrow$
       & $\downarrow$	& $\downarrow$	& $\uparrow$	& $\downarrow$	& $\uparrow$	& $\updownarrow$
       & $\uparrow$	& $\uparrow$	& $\uparrow$	& $\uparrow$	& $\uparrow$	& $\downarrow$ \\
   \toprule
  \end{tabular}   
  \caption{Changes in the DDB soliton characteristics upon considering different variations of the systems' parameters 
  and monitoring the four right moving DDB solitons generated for fixed $|u_{+1}|=1$ and $a_{+1}=a_0=5$.
  Here, $(1)$ [$(4)$] refers to the inner [outer] most DDB state [see Figs.~\ref{fig:DDB}(b) and \ref{fig:DDB}(e)]. 
   The top row indicates the distinct variations, namely $|u_0|$, $A_{-1}$, and $\kappa_{-1}$, 
  performed within the interval $[0.5, 2]$. 
  The second row contains the soliton characteristics, i.e. the dark, $\nu_{+1,0}$, and bright, $\eta_{-1}$, amplitudes,  
  the common inverse width, $\mathcal{D}$, the normalized number of particles, $n_b$, 
  hosted in the bright soliton component and the velocity, $\dot{x}_0$, of the DDB pair. 
  $\uparrow$ arrows ($\downarrow$) indicate an increase (decrease) of the corresponding quantity. 
  $\updownarrow$ arrows indicate that within the above interval a non-monotonic tendency of the respective quantity is 
  observed.}
  \label{tab:DDB_char}
 \end{table*}  

The same overall picture is qualitatively valid for the corresponding DBB soliton formation.
Note that in contrast to the DDB nucleation,
to generate DBB soliton arrays the counterflow dynamics is featured solely by one of the hyperfine components 
which as per our choice is the $m_{F}={+1}$ one.  
The remaining two hyperfine components, namely $m_{F}={0, -1}$ 
share the same Gaussian-shaped initial profile.
In Figs.~\ref{fig:DBB}(a)-\ref{fig:DBB}(f) the formation of four, six and eight DBB soliton
complexes is shown for $a_{+1}=3$, $a_{+1}=5$ and $a_{+1}=7$ respectively.
Notice that the number of the generated DBB states appears
to be lower when compared to the DDB solitons formed for the same value of $a_{+1}$.
For instance, four DBB solitons are formed for $a_{+1}=3$ [Figs.~\ref{fig:DBB}(a), \ref{fig:DBB}(d)] while the corresponding 
DDB soliton count is six [Figs.~\ref{fig:DDB}(a), \ref{fig:DDB}(d)].
The observed difference between the number of nucleated DBB and DDB states can be intuitively 
understood as follows. 
For a DDB production the total number of particles is $N=2990$ while for a DBB one is $N=1498$ 
e.g. for the case examples presented in 
Figs.~\ref{fig:DDB}(a), \ref{fig:DDB}(d) and Figs.~\ref{fig:DBB}(a), \ref{fig:DBB}(d) respectively.
Recall that in our simulations we fix the chemical potential and thus $N$
is a free parameter. 
The significantly lower number of particles in a DBB nucleation process stems from the fact 
that two of the participating components have a Gaussian initial profile and as such host fewer particles.
This decrease of the system size for a DBB realization when compared to a DDB one 
may be partially responsible for the observed 
decreased DBB soliton count. 
Moreover, in the DDB case the presence of two components (namely $m_F=+1, 0$ components each one characterized by an 
amplitude $|u|$) with a finite background leads to a total amplitude  $|u_{eff}|\approx 2|u|$.
Thus, as dictated by Eq. (7), the number of solitons is expected to be higher as well. 
Further adding to the above, for a DBB formation only one component develops, via interference, dark solitons. 
These dark solitons are, in turn, responsible for the trapping 
of bright solitons in the other two hyperfine components.
However, since two components develop bright solitons effectively the number of particles 
that have to be sustained by each effective dark well increases. 
As such in the DBB case, the system prefers to develop fewer but also wider and deeper dark solitons than in the DDB process; 
this is also inter-related with the smaller counterflow induced momentum in the DBB cse. 
These deeper dark solitons can in turn efficiently trap and waveguide
the resulting also fewer bright solitons. 
The above intuitive explanation is fairly supported by our findings.
Indeed, both the dark and the bright solitons illustrated in 
Figs.~\ref{fig:DBB}(a), \ref{fig:DBB}(d) appear to be wider having also larger amplitudes 
when compared to the ones formed in the DDB interference process shown in Figs.~\ref{fig:DDB}(a), \ref{fig:DDB}(d).

In all cases presented in Figs.~\ref{fig:DDB}(a)-\ref{fig:DDB}(f) 
and Figs.~\ref{fig:DBB}(a)-\ref{fig:DBB}(f), we were able to showcase upon fitting that 
the evolved dark and bright states have the standard $\tanh$- and $\sech$-shaped waveform 
respectively [see Eqs.~(\ref{eq:DB_d})-(\ref{eq:DB_b})]. 
Moreover, following the procedure described in the two-component setting (see Sec.~\ref{subsec:psuedo-spinor}),
we verified that the number of particles hosted in each of the bright solitons formed follows Eq.~(\ref{eq:DB_N}),
with the common inverse width, $\mathcal{D}$, satisfying the generalized conditions    
\begin{eqnarray}
\mathcal{D}^2&=&\nu_j^2+\nu_k^2-\eta_l^2, \label{eq:DDB_D}
\\
\mathcal{D}^2&=&\nu_j^2-\eta_k^2-\eta_l^2, \label{eq:DBB_D}
\end{eqnarray}
for the DDB and the DBB cases respectively. 
The indices $j,k,l$ in the above expressions denote the three (distinct) hyperfine 
components.
As a case example, for one of the DDB states
shown in Figs.~\ref{fig:DDB}(b), \ref{fig:DDB}(e) $N^{num}_b=0.3715$ while the semi-analytical 
prediction gives $N_b=0.3721$. Notice that the deviation is again smaller than $1\%$.

As a next step, we attempt to appreciate the effect that  
different initial configurations have on the characteristics of the resulting DDB and DBB soliton compounds.
Our findings are summarized in Tables~\ref{tab:DDB_char} and \ref{tab:DBB_char} respectively.
Specifically, for a DDB nucleation process $|u_{+1}|=1$ and $a_{+1}=a_0=5$ are held fixed.
The remaining parameters are varied (one of them at a time) within the interval $[0.5, 2]$.
The above selection leads to the appearance of eight DDB solitons symmetrically formed around the 
origin as already illustrated in Figs.~\ref{fig:DDB}(d), \ref{fig:DDB}(e). 
Exploiting this symmetric formation only the four, i.e. (1)-(4), right moving DDB solitons
indicated in Figs.~\ref{fig:DDB}(d), \ref{fig:DDB}(e) are monitored and shown in Table~\ref{tab:DDB_char}. 
\begin{table*}  
  \def\arraystretch{1.5}
  \centering  
  \begin{tabular}{c | ccccccc | ccccccc | ccccccc} 
  \toprule
  [0.5 , 2]
       & \multicolumn{7}{c}{$\abs{u_{+1}}  \uparrow$}
       & \multicolumn{7}{ | c | }{$A_{0}  \uparrow$}
       & \multicolumn{7}{c}{$\kappa_{0} \uparrow$} \\
   \toprule
   DDB
       & $\nu_{+1}$	& $\eta_0$	& $\eta_{-1}$	& $\mathcal{D}$	& $n_0$ 	& $n_{-1}$	& $\dot{x}_0$
       & $\nu_{+1}$	& $\eta_0$	& $\eta_{-1}$	& $\mathcal{D}$	& $n_0$ 	& $n_{-1}$	& $\dot{x}_0$
       & $\nu_{+1}$	& $\eta_0$	& $\eta_{-1}$	& $\mathcal{D}$	& $n_0$ 	& $n_{-1}$	& $\dot{x}_0$ \\ 
   \hline
   (1) & $\uparrow$	& $\uparrow$	& $\uparrow$	& $\uparrow$	& $\uparrow$	& $\uparrow$	& $\uparrow$
       & $\downarrow$	& $\uparrow$	& $\downarrow$	& $\downarrow$	& $\downarrow$	& $\downarrow$	& $\downarrow$
       & $\uparrow$	& $\downarrow$	& $\uparrow$	& $\uparrow$	& $\downarrow$	& $\downarrow$ 	& $\updownarrow$\\
   (2) & $\uparrow$	& $\uparrow$	& $\uparrow$	& $\uparrow$	& $\uparrow$	& $\uparrow$	& $\uparrow$
       & $\downarrow$	& $\uparrow$	& $\downarrow$	& $\downarrow$	& $\downarrow$	& $\downarrow$	& $\downarrow$
       & $\uparrow$	& $\downarrow$	& $\uparrow$	& $\uparrow$	& $\updownarrow$& $\uparrow$ 	& $\updownarrow$\\
   (3) & $\uparrow$	& $\uparrow$	& $\uparrow$	& $\uparrow$	& $\downarrow$	& $\downarrow$	& $\uparrow$
       & $\downarrow$	& $\uparrow$	& $\downarrow$	& $\downarrow$	& $\uparrow$	& $\uparrow$	& $\downarrow$
       & $\uparrow$	& $\downarrow$	& $\uparrow$	& $\uparrow$	& $\updownarrow$& $\updownarrow$& $\updownarrow$ \\
   \toprule
  \end{tabular}   
  \caption{Same as Table~\ref{tab:DDB_char} but for the three right moving 
    DBB solitons generated for fixed $a_{+1}=5$. $(1)$ [$(3)$] denotes the
    inner [outer] most DBB structure shown in Figs.~\ref{fig:DBB}(b) and \ref{fig:DBB}(e).
    Other parameters used are $A_{-1}=\kappa_{-1}=1$.}
  \label{tab:DBB_char}
 \end{table*}  

It becomes apparent, by comparing the DDB results of Table~\ref{tab:DDB_char} 
with the ones found in the two-component scenario (see Table~\ref{tab:DB_char}),
that the inclusion of an extra $m_F=+1$ component prone to host dark solitons
leads to the following comparison of the resulting 
soliton characteristics.
As $|u_0|$ is increased within the interval $[0.5, 2]$, 
all the generated DDB states appear to be faster and narrower 
(see second column in Table~\ref{tab:DDB_char}).
Additionally, larger amplitudes are observed 
for all the solitons in all the three hyperfine components.
The same qualitative results were also found in the relevant variation but for the two-component system
(see second column in Table~\ref{tab:DB_char}).
It is important to stress at this point that further increase of $|u_0|$ 
and/or different initial values of $|u_{+1}|$,
can lead to a change in the number of states generated as suggested
also by Eq.~(\ref{eq:w_n_IP}).
This is the reason for limiting our variations to the aforementioned interval.
However, our findings for choices of $|u_{+1}| \neq |u_{0}|$ suggest that, 
given the same half-width $a_{+1}=a_{0}$, the number of 
nucleated DDB solitons will be determined by the larger $|u_{m_F}|$ value.

On the contrary, upon varying the amplitude, $A_{-1}$,  
of the initial Gaussian pulse the impact of the additional $m_F=+1$ component   
is imprinted on the velocity outcome of the resulting DDB solitons 
(see third column in Table~\ref{tab:DDB_char}). 
Indeed, as $A_{-1}$ increases a uniquely defined tendency of the velocity of the resulting 
states cannot be inferred at least within the interval of interest here. 
This result differs from the systematic
overall decrease of the DB soliton velocity 
observed in the two-component scenario (see third column in Table~\ref{tab:DB_char}).
Notice also that all the remaining soliton characteristics here are similar to the ones found in the two-component scenario
(compare the third column in Tables~\ref{tab:DDB_char} and~\ref{tab:DB_char}, respectively).
Additionally, the presence of the extra $m_F=+1$ 
component leads to no modification
on the observed DDB soliton characteristics 
when considering variations of the 
inverse width, $\kappa_{-1}$, of the Gaussian pulse (see fourth column in Table~\ref{tab:DDB_char}). 
Namely, for increasing $\kappa_{-1}$ all the resulting DDB states are narrower and slower.
An outcome that was also found in this    
type of variation but in the two-component setting (see fourth column in Table~\ref{tab:DB_char}).

Next we will check the same diagnostics but for the DBB nucleation process.
Along the same lines, the initial parameters used for a DBB realization are
$A_0=\kappa_0=1$ and $a_{+1}=5$ [as per Eq.~(\ref{eq:gaussian}) and Eq.~(\ref{eq:square_well}) respectively]. 
This choice, results in the six DBB solitons illustrated in Figs.~\ref{fig:DBB}(d), \ref{fig:DBB}(e). 
Again due to symmetry only the three, (1)-(3), right moving states indicated in Figs.~\ref{fig:DBB}(d), \ref{fig:DBB}(e)
are monitored in Table~\ref{tab:DBB_char}. 
When comparing the relevant findings presented in Table~\ref{tab:DBB_char} to those shown in 
Table~\ref{tab:DB_char} the following conclusions can be drawn. 
For increasing $|u_{+1}|$ the generated DBB solitons are found to be narrower and faster 
similarly to the evolved DB states observed in the two-component scenario
(see second column in Table~\ref{tab:DB_char}).
Alterations occur only upon varying the characteristics of the initial Gaussian pulse. 
In particular, the effect of adding an extra bright component upon increasing the amplitude, $A_{0}$, 
of the initial Gaussian pulse is the observed increased amplitude of all bright solitons formed in this component
(see third column in Table~\ref{tab:DBB_char}).
Yet, all the resulting DBB states are found to be wider and slower for increasing $A_{0}$,
an outcome which is similar to that found in the
two-component setting (see third column in Table~\ref{tab:DB_char}).
Lastly, upon increasing $\kappa_{0}$ the impact that the extra 
$m_F=0$ component has on the resulting DBB solitons is the following (see fourth column in Table~\ref{tab:DBB_char}). 
Besides the observed decreased amplitude of all bright solitons formed
in this component, the outermost DBB states, i.e. (2), (3), are the ones that are affected the most. 
Notice that a non-monotonic response of the normalized number of particles, $n_{0}$, 
hosted in this $m_F=0$ component is found as $\kappa_{0}$ increases within the interval $[0.5, 2]$. 
Additionally, the velocity of all three DBB solitons shows a non-monotonic tendency as $\kappa_{0}$ is increased. 
It is relevant to note that this result is in contrast to the decrease observed in the two-component scenario 
(see fourth column in Table~\ref{tab:DB_char}).

It is worth commenting at this point that in both of the above-discussed processes 
we also considered variations of the relevant in each case $a_{m_F}$.
Recall that $a_{+1,0}$ and $a_{+1}$ are the associated half-widths of the initial IP-IRP 
for a DDB and DBB nucleation process respectively.
In particular, by fixing all parameters to their default values (see here Sec.~\ref{subsec:setup_homo})
we varied the relevant $a_{m_F}$ within the interval $[1, 10]$.
The general conclusion for such a variation, in both processes, is that increasing $a_{m_F}$ results to
more states which become narrower and slower as their number is increased. 
Differences here, mostly refer to the relevant amplitudes
of the resulting solitons and the normalized number of particles hosted in each bright soliton constituent.
Importantly, and referring solely to the DDB process, it is found that 
given the same initial amplitude, $|u_{m{_F}}|$, the number of solitons generated depends on the 
smallest initial $a_{m_F}$. For this latter case ($a_{+1} \neq a_{0}$)
a spatially modulated density background occurs for the components hosting the dark states.
Finally, and also in both processes  
we were able to verify that for displacements, $X_0$, of the initial Gaussian pulse 
$X_0\geq a$ generation of DDB and DBB solitons is absent.
This is in line with our findings in the two-component system. 
 \begin{figure}[t]
 \centering
 \includegraphics[width=0.48\textwidth]{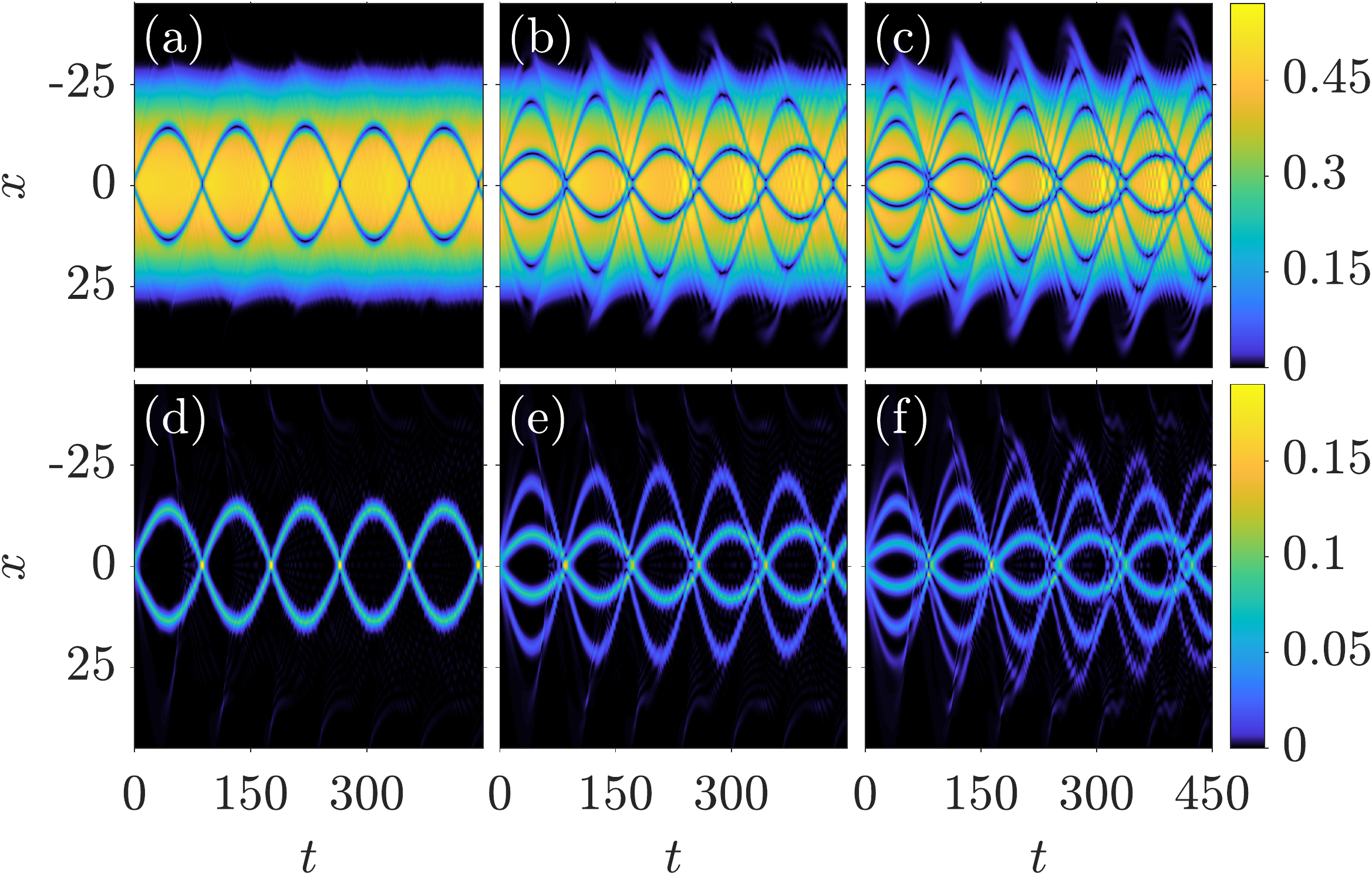}
 \caption{Evolution of the densities $|\Psi_{+1}|^2$, $|\Psi_{-1}|^2$ showcasing the generated 
     DDB solitons in a harmonic trap with $\Omega=0.05$.
 	 Increasing the width $w$ of the double-well barrier 
 	 allows the generation of DDB soliton arrays consisting of two [(a), (b)], four [(c), (d)] and six [(e), (f)]
 	 DDB solitons respectively for $w^2=1$, $w^2=5$ and $w^2=10$.
 	 In all cases top (bottom) panels illustrate the formation of dark (bright) solitons in the $m_F=+1$ ($m_F=-1$) component.
 	 Since the evolution of the $m_F=0$ component is identical to the one shown for the $m_F=+1$ component it is omitted.
 	 }
 \label{fig:DDB_trap} 
 \end{figure}

We now turn to the trapped three-component DDB and DBB case, in analogy
with the corresponding two-component one.
As in the latter, for the systematic production of multiple DDB and DBB solitons 
we use as a control parameter the width, $w$, of the double-well potential [see Eq.~(\ref{eq:V_m})].
Figures~\ref{fig:DDB_trap}(a)-\ref{fig:DDB_trap}(f) illustrate the formation of two, four and six DDB
solitons for $w^2=1$, $w^2=5$ and $w^2=10$ respectively. 
In all cases presented in this figure top (bottom) panels depict the evolution of the density, $|\Psi_{+1}|^2$, 
($|\Psi_{-1}|^2$) of the $m_F=+1$ ($m_F=-1$) component. 
Notice the close resemblance of the dynamical evolution of the DDB states when compared to the relevant evolution 
of the DB soliton arrays shown in Figs.~\ref{fig:DB_trap}(a)-\ref{fig:DB_trap}(f). 
We remark here, that the above-observed evolution holds equally for the corresponding DBB states (results not shown here for brevity). 
\begin{figure}[t]
\centering
\includegraphics[width=0.48\textwidth]{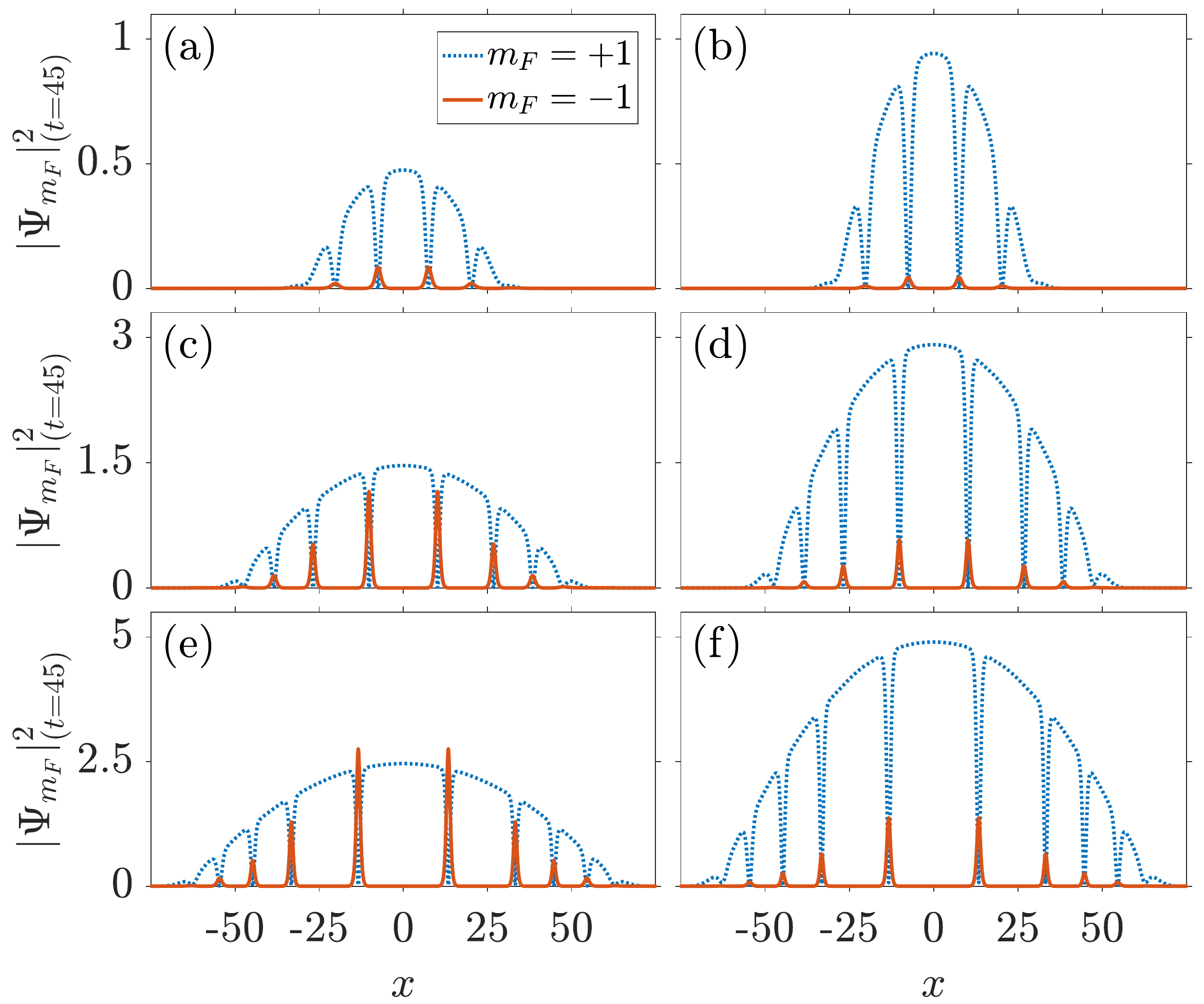}
\caption{Profile snapshots of the density $\abs{\Psi_{m_{F}}}^2$ with $m_F=\pm 1$
at $t=45$, illustrating the generated DDB (left) and DBB (right) solitons in the trapped scenario. 
In all cases $w^2=5$ and we vary the corresponding chemical potential.
From top to bottom $\mu=1$, $\mu=3$ and $\mu=5$, resulting in turn to 
four [(a)-(b)], six [(c)-(d)], and eight [(e)-(f)] DDB-DBB solitons respectively.  
The $m_F=0$ component is omitted since it shows the same profile as the $m_F=+1$ ($m_F=-1$) for the DDB (DBB)
nucleation process.}
\label{fig:DDB_vs_DBB_trap} 
\end{figure}

Furthermore, below we briefly report on the systematic production of the desired number of DDB and 
DBB solitons upon varying the common chemical potential $\mu$ of the confined three-component system.
In Figs.~\ref{fig:DDB_vs_DBB_trap}(a)-\ref{fig:DDB_vs_DBB_trap}(f)
a direct comparison of the resulting DDB and DBB soliton compounds is provided for three
different values of $\mu$. 
Evidently, the number of DDB and DBB soliton complexes generated in each different initialization 
is exactly the same and as expected, it increases for increasing $\mu$.
E.g for $\mu=3$ illustrated in Figs.~\ref{fig:DDB_vs_DBB_trap}(c) and \ref{fig:DDB_vs_DBB_trap}(d)   
for the DDB and DBB processes respectively, the nucleated states at $t=45$ are six. 
Note also, that in all cases the $m_F=0$ component overlaps either with $m_F=+1$ component (DDB nucleation process) 
or with the $m_F=-1$ (DBB nucleation process) and as such it is not shown in the relevant profiles. 
Concluding, the dynamical evolution of both types of soliton arrays is qualitatively the same and closely resembles the 
one observed in the two-component setting. Also, in all the different parametric variations and for both nucleation 
processes studied above, the resulting arrays of DDB and DBB solitons remain robust, while oscillating and 
colliding with one another, for evolution times up to $t=450$ that we have checked.

\subsection{Spinor BEC}\label{subsec:spinor}

\begin{figure}[t]
\centering
\includegraphics[width=0.48\textwidth]{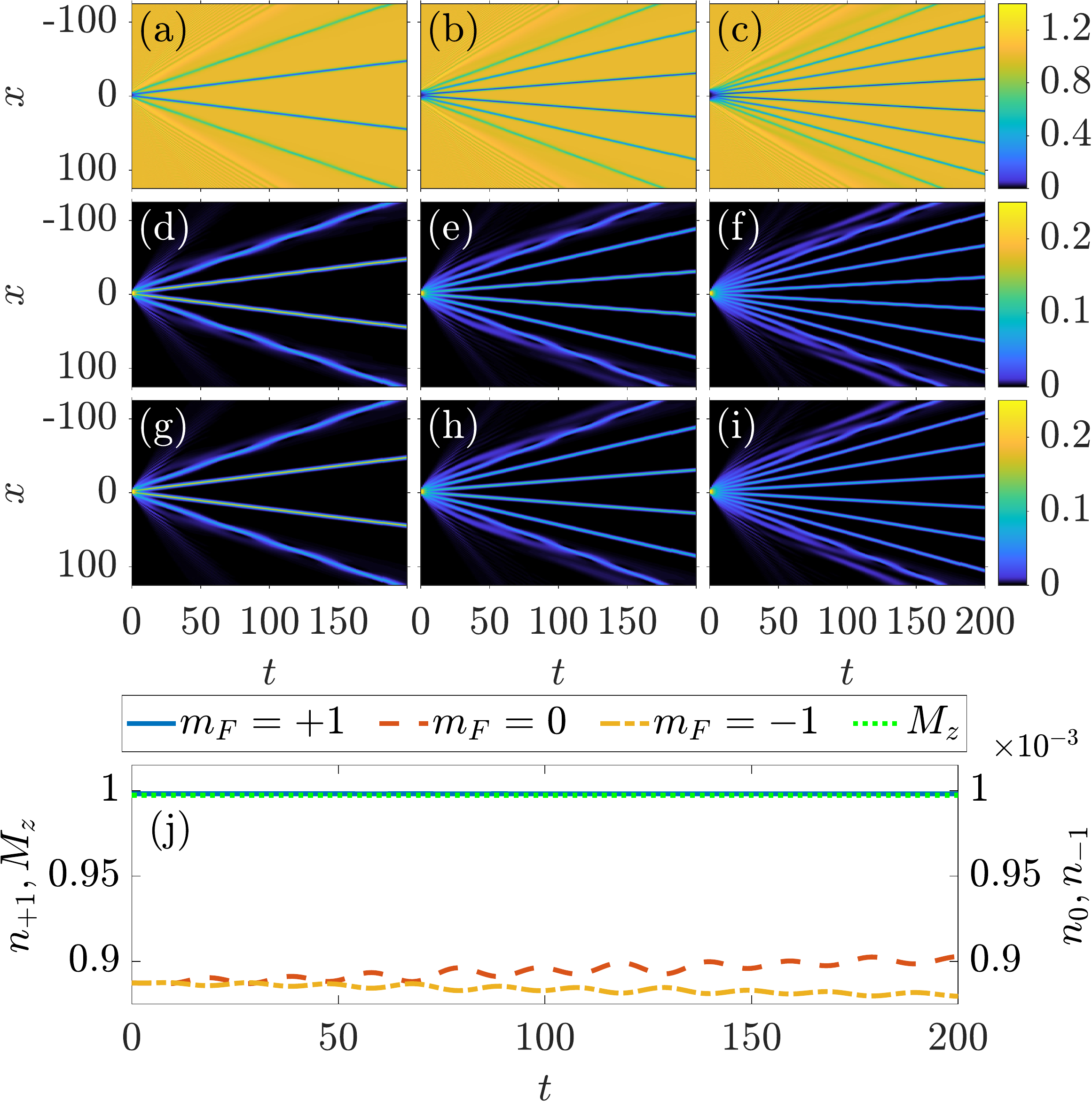}
\caption{Spatio-temporal evolution of the density $\abs{\Psi_{m_F}}^2$ 
of the (a)-(c) $m_F={+1}$, (d)-(f) $m_F={0}$, and (g)-(i) $m_F={-1}$ 
component respectively for varying $a_{+1}$. 
From left to right $a_{+1}=3$, $a_{+1}=5$ and $a_{+1}=7$, 
allowing the generation of four [(a), (d), (g)], six [(b), (e), (h)] and 
eight [(c), (f), (i)] DBB solitons respectively for the homogeneous spinor setting.
In all cases, (a)-(c) illustrate the formation of dark solitons in the $m_F={+1}$ component and (d)-(f) [(g)-(i)] 
depict the formation of bright solitons in the $m_F={0}$ [$m_F={-1}$] component of the spinor system.
(j) Evolution of the population, $n_{m{_F}}$, of each hyperfine component, 
and of the total magnetization, $M_z(t)$ for $a_{+1}=5$.
Notice that $n_0$ and $n_{-1}$ are three orders of magnitude smaller than $n_{+1}$ which is why 
a second axis is introduced.}
\label{fig:SDBB} 
\end{figure}

Up to now, the controlled formation of multiple soliton complexes of the DB type in two- and three-component BECs
has been established. In what follows we turn our attention to the spinor $F=1$ BEC~\cite{Bersano2018}.
In this way, we will be able to address the fate of the generated DDB and DBB soliton arrays
when spin degrees of freedom are taken into account.
Recall, that the evolution of this system is dictated by 
Eqs.~(\ref{eq:spinor_hamiltonian_a})-(\ref{eq:spinor_hamiltonian_b}). 
In order to induce the dynamics we will utilize once more the counterflow processes introduced in 
Sec.~\ref{subsec:setup_homo}.

\begin{figure}[t]
\centering
\includegraphics[width=0.48\textwidth]{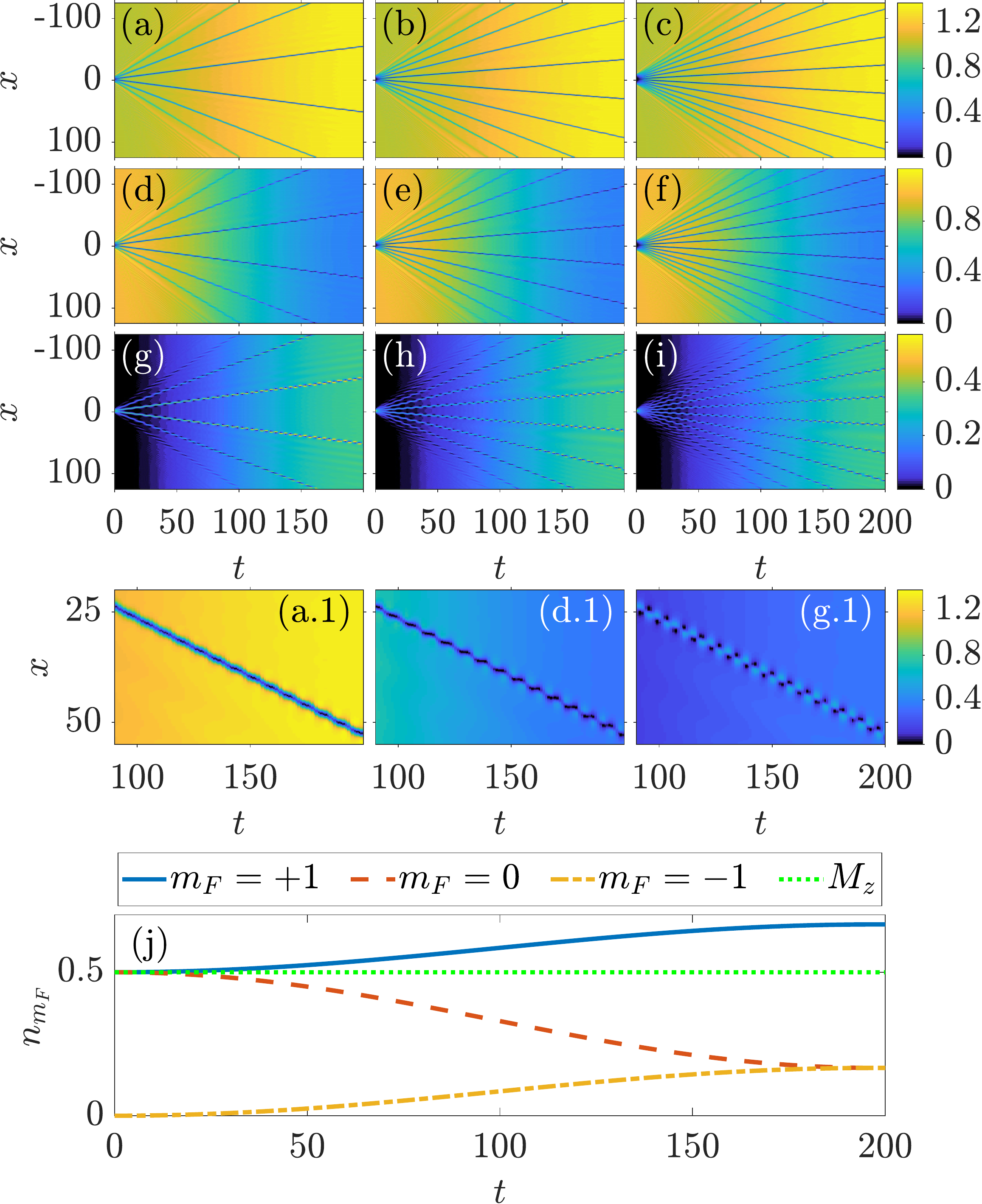}
\caption{ Same as in Fig.~\ref{fig:SDBB} but showcasing the generation of DDB solitons. 
In all cases, (a)-(c) [(d)-(f)] illustrate the formation of dark solitons in the $m_F={+1,0}$ components and 
(g)-(i) illustrate the formation of bright solitons in the $m_F={-1}$ component of the spinor system. 
($a_1$), ($d_1$), ($g_1$) Trajectory of the closest to the origin right moving originally formed DDB soliton shown in (a), (d), (g)
transitioning during evolution into beating dark states.
(j) Evolution of the population, $n_{m_F}$, of each hyperfine component, 
as well as of the total magnetization, $M_z(t)$, for $a_{+1}=a_{0}=5$.}
\label{fig:SDDB} 
\end{figure}

As usual, we start our analysis by considering the homogeneous system. 
As in the previous section, for a DDB generation process the initial ansatz used 
for the $m_F=+1,0$ [$m_F=-1$] components is given by Eq.~(\ref{eq:square_well}) [Eq.~(\ref{eq:gaussian})].
Accordingly, to dynamically produce DBB soliton arrays the corresponding initial conditions  
are provided by Eq.~(\ref{eq:gaussian}) [Eq.~(\ref{eq:square_well})] for the $m_F=0,-1$ [$m_F=+1$] hyperfine components.  
Figures~\ref{fig:SDBB}(a)-\ref{fig:SDBB}(j) and Figs.~\ref{fig:SDDB}(a)-\ref{fig:SDDB}(j) 
summarize our numerical findings. 
In particular, Figs.~\ref{fig:SDBB}(a)-(i) illustrate the evolution of the density, $\abs{\Psi_{m_F}}^2$, 
of all three $m_F=+1,0,-1$ components.
The controlled generation 
of four, six and eight DBB soliton arrays can be readily seen as $a_{+1}$ is increased from $a_{+1}=3$ to $a_{+1}=7$ 
[see Figs.~\ref{fig:SDBB}(a), (d), (g), Figs.~\ref{fig:SDBB}(b), (e), (h) and Figs.~\ref{fig:SDBB}(c), (f), (i), 
respectively]. 
Comparing the dynamical evolution of the spinor system to the one observed in the corresponding three-component setting
[see Figs.~\ref{fig:DBB}(a)-\ref{fig:DBB}(f)]
it becomes apparent that the inclusion of the spin interaction
has a minuscule effect on both the nucleation and the long time evolution of the DBB states.  
To appreciate the latter, in Fig.~\ref{fig:SDBB}(j) we monitor the temporal evolution of the 
population, i.e. $n_{m_{F}}(t)=\frac{1}{N}\int |\Psi_{m_{F}}\left(x,t\right)|^2\text{d}x$, 
of each hyperfine component, as well as the total magnetization 
$M_z(t)=\int\left( |\Psi_{+1}\left(x,t\right)|^2-|\Psi_{-1}\left(x,t\right)|^2\right)\text{d}x$ 
of the spinor system for $a_{+1}=5$ (see also Sec.~\ref{subsec:models}).
Note that $n_{0}(t)$, $n_{-1}(t)$ are multiplied by a factor of $10^3$ in order to be visible,
and that the same picture holds equally for all the distinct variations of $a_{+1}$ presented in Fig.~\ref{fig:SDBB}.
As it can be deduced, oscillations of $n_{+1}(t)$, $n_{0}(t)$ and $n_{-1}(t)$ occur during the evolution.
Recall now, that a spinor condensate is subject to the so-called spin relaxation process. 
The latter, allows for collisions of two $m_F=0$ atoms that can in turn produce a pair of particles in the 
$m_F=+1$ and $m_F=-1$ component and vice versa~\cite{Chang2005}.
It is this continuous exchange of particles
that leads to the oscillatory trajectories observed for the bright soliton constituents of the resulting DBB arrays. 
Notice that $n_{+1}(t)$ is significantly larger when compared to $n_{0,-1}(t)\sim 10^{-3}$ 
and due to the rescaling used appears almost constant ($n_{+1}(t)\approx1$) during evolution.  
However, we must stress that also not discernible    
oscillations of the population of this hyperfine component are present and are similar to the ones observed for 
the $n_{-1}(t)$ component. Therefore $M_z(t)$ remains constant during
the evolution while being of order unity.

Contrary to the DBB nucleation process investigated above, for a DDB realization  
the spin-mixing dynamics plays a crucial role.
As in the previous scenario, Figs.~\ref{fig:SDDB}(a)-\ref{fig:SDDB}(i) show 
the spatio-temporal evolution of the densities, $\abs{\Psi_{m_F}}^2$, of all three $m_F$ components. 
Also here, by manipulating $a_{+1}=a_{0}=a$ we were able to controllably generate
arrays of DDB solitons in this homogeneous spinor setting. 
From left to right in this figure, six, eight and twelve solitons are formed, 
corresponding to $a=3$, $a=5$ and $a=7$ respectively.
In particular, Figs.~\ref{fig:SDDB}(a)-(c) [Figs.~\ref{fig:SDDB}(d)-(f)] depict the dark solitons formed 
in the $m_F=+1$ [$m_F=0$] component.
Additionally, Figs.~\ref{fig:SDDB}(g)-(i) illustrate the bright states formed in the respective $m_F=-1$ component.
Strikingly enough, as it is observed in all of the aforementioned contour plots, 
as time evolves, the background density gradually changes (notice the change in the color gradient).
This result, as we will show later on, is attributed to the spin-mixing dynamics
that significantly alters the evolution of the DDB
soliton arrays formed. 

To shed light on the observed dynamics, below
let us focus our attention on Figs.~\ref{fig:SDDB}(a), \ref{fig:SDDB}(d) and \ref{fig:SDDB}(g) for $a=3$. 
Here, also a zoom is provided in Figs.~\ref{fig:SDDB}($a.1$), \ref{fig:SDDB}($d.1$) and \ref{fig:SDDB}($g.1$)
to elucidate our analysis. In the latter figures
the closest to the origin DDB pair is monitored. 
As time evolves the background density of the $m_F=+1$ component increases which suggests that transfer 
of particles from the lower hyperfine components takes place.
The latter can indeed be confirmed by inspecting the evolution of the $m_F=0$ component. 
Evidently, the background density of this component gradually decreases.
The corresponding density of the $m_F=-1$ component is also seen to increase.   
Monitoring the evolution of the respective populations, $n_{m_{F}}(t)$, shown in Fig.~\ref{fig:SDDB}(j),
delineates the above trend. 
Indeed, at $t=0$ $n_{+1}(0)=n_{0}(0)=0.5$ while $n_{-1}(0)\sim 10^{-3}$.
However, during evolution $n_{+1}(t)$ increases reaching the value of  $n_{+1}(t=200)\approx 0.66$. 
Accordingly, $n_{0}(t)$ decreases drastically during propagation acquiring a similar value with 
$n_{-1}(t)$ at later evolution times, i.e. $n_{0}(t=200) \approx n_{-1}(t=200)\approx 0.16$. 
Note also, that the total magnetization of the system is preserved with $M_z(t)=0.5$
throughout the evolution.
Returning now to the relevant densities, since the background density of the $m_F=0$ component  
decreases, the dark states formed in this component begin to deform.
At later times ($t>150$) the solitonic states developed in this hyperfine
component have both a dark and a bright component [see Fig.~\ref{fig:SDDB}($d.1$)].
Similarly, at early times the $m_F=-1$ component hosts bright solitons.
Since the number of particles, in this case increases, a finite background slowly appears~\cite{Li2005,Kurosaki2007}.
As such, also the bright solitons of this component begin to deform. 
The latter deformation leads in turn to the formation of solitonic structures that 
again have both a bright and a dark part, involving a breathing
between the two and are formed also faster in this $m_F=-1$
when compared to $m_F=0$ one [see Fig.~\ref{fig:SDDB}($g.1$)].  
The same deformation occurs also in the $m_F=+1$ component but at propagation times even larger than the 
ones depicted in Fig.~\ref{fig:SDDB}(a). Indeed, by inspecting the evolution
of the closest to the origin dark soliton of the originally formed DDB state shown in Fig.~\ref{fig:SDDB}($a.1$),
the dark soliton is also deformed in this case, yet the beating pattern
of panel ($g.1$) [and even that of ($d.1$)] is not as straightforwardly
discernible.
Nevertheless, close inspection indicates
that the evolved states in all three $m_F$ components bear similar characteristics
to the so-called beating dark solitons that were experimentally observed in
two-component systems~\cite{Yan2012}.
As such, these states can be thought of as the generalization of the beating
dark solitons in spinor BECs.

Before proceeding to the harmonically confined spinor BEC system, a final 
comment is of relevance here. 
Investigating the current setting, we also considered different initializations
in which the symmetric, with respect to the $m_F=0$, hyperfine states have the same initial 
conditions. In this way we were able to generate symmetric variants of the DDB and
DBB states discussed above. Namely, DBD and BDB soliton arrays.
In these cases, our simulations indicate that the resulting states show all features found in the three-component setting.
Although the spin interaction is present, the conversion of particles from one component to another is six orders of 
magnitude smaller than the total number of particles.
As such the spin interaction is negligible. For these systems, also the total magnetization is
zero, in contrast to the finite one observed for the asymmetric, in the above sense, DDB and DBB 
soliton arrays addressed herein. In that light, it appears as if the
drastic effect of the spin-interaction contribution in the previous
realization is able to excite the beating dark soliton generalizations.
On the other hand, following the approach of~\cite{Yan2012} in the
three-component Manakov case, it is also possible to excite beating solitons
in the latter (spin-independent) case. However, a more systematic theoretical
analysis of the beating states is deferred for a separated work.

\begin{figure}[t]
\centering
\includegraphics[width=0.48\textwidth]{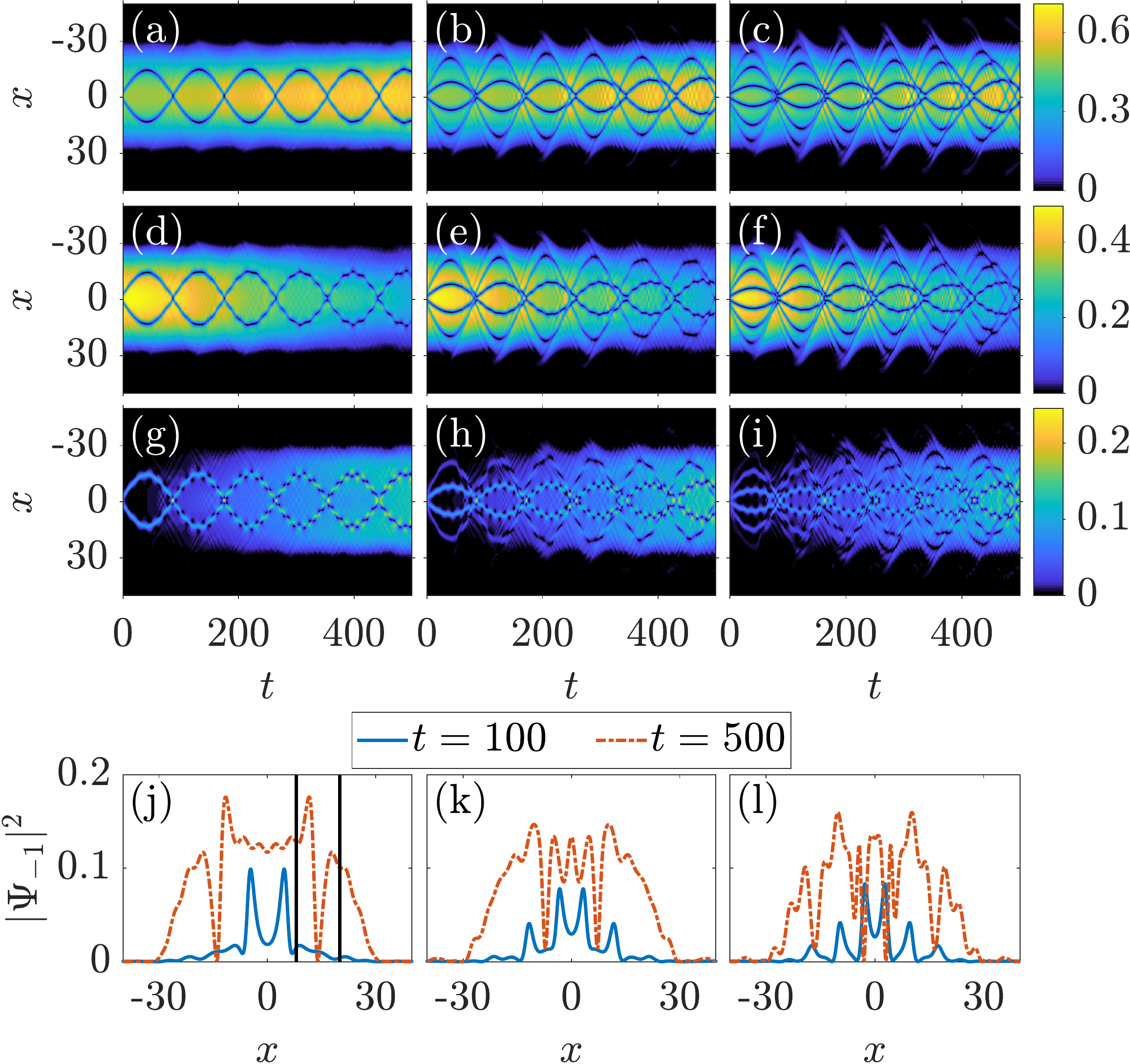}
\caption{
Spatio-temporal evolution of the densities $\abs{\Psi_{m_F}}^2$ of the (a)-(c) $m_F={+1}$, (d)-(f) $m_F={0}$ and (g)-(i) 
$m_F={-1}$ components upon varying the width, $w$, of the double-well barrier.
From left to right $w^2=1$, $w^2=5$ and $w^2=10$, allowing the generation of two [(a), (d), (g)], four [(b), (e), (h)]   
and six [(c), (f), (i)] DDB solitons respectively in the spinor system.
In all cases, (a)-(c) [(d)-(f)] illustrate the formation of dark solitons 
in the $m_F={+1}$ [$m_F={0}$] component and (g)-(i) 
the generated bright solitons, transitioning into beating dark states, in the $m_F={-1}$ component.
(j)-(l) Vertical cuts of $\abs{\Psi_{-1}}^2$ for the three distinct values of $w$ (see legend).
In (j) solid rectangle indicates a beating dark soliton.}
\label{fig:SDDB_trap}
\end{figure} 
In the trapped scenario we exclusively present our findings for a DDB generation process. 
This is due to the fact that only this nucleation process entails new features stemming from the spin-mixing dynamics.
The initial state preparation used herein is the one described in the confined three-component setting 
(see Sec.~\ref{subsec:three-component}).
Once more, by properly adjusting the initial width, $w$, of the double-well potential [see Eq.~(\ref{eq:V_m})]
the controlled formation of multiple DDB soliton complexes is achieved in this harmonically trapped spinor system.   
From top to bottom Figs.~\ref{fig:SDDB_trap}(a)-(i) 
show the evolution of $\abs{\Psi_{m_F}}^2$ (with $m_F=+1,0,-1$).  
As $w^2$ is increased from $w^2=1$ to $w^2=10$ two, four and six such
solitons are formed 
[e.g. see Figs.~\ref{fig:SDDB_trap}(a)-(c)].
Dark solitons emerge in the $m_F=+1,0$ components [see Figs.~\ref{fig:SDDB_trap}(a)-(c), \ref{fig:SDDB_trap}(d)-(f)]
while bright states are generated in 
the $m_F=-1$ component [see Figs.~\ref{fig:SDDB_trap}(g)-(i)].
However, as in the homogeneous scenario, soon after their formation
all states formed and also in all hyperfine components begin to deform.
This deformation occurs faster in the less populated $m_F=-1$ component 
and later on in the other two hyperfine states. 
This phenomenon is yet again attributed to the spin-mixing dynamics
that allows for particle exchange between the components.
Focusing  on Figs.~\ref{fig:SDDB_trap}(a), (d), (g), 
the background densities of both the $m_F=\pm1$ components increase 
while the density of the $m_F=0$ one decreases.
This exchange in population leads in turn during evolution to a transition 
of the soliton states in each component into states that
bear both a dark and a bright part.
Thus, in line with our findings in the homogeneous case,
beating dark solitons are progressively formed in all three hyperfine components.
Since these beating structures are more pronounced in the $m_F=-1$ component,
in Fig.~\ref{fig:SDDB_trap}(j) profile snapshots of the density of this component 
are illustrated. In particular, $|\Psi_{-1}|^2$ is depicted for  
two different time instants, namely $t=100$ and $t=500$, during
the evolution and for $w^2=1$.
At initial times the two bright solitons originally formed in this component
are now on top of a still small, yet finite background.
Namely, they are already deformed into states that are reminiscent of the
so-called antidark 
solitons~\cite{danaila2016vector,kevrekidis2004families,Katsimiga2017}. 
At larger evolution times, instead of the aforementioned antidark solitons, 
two beating dark states are seen to propagate. One of them is indicated
in Fig.~\ref{fig:SDDB_trap}(j) by a black rectangle.  
Notice that this beating state has a density dip followed be a density hump.

\begin{figure}[t]
\centering
\includegraphics[width=0.48\textwidth]{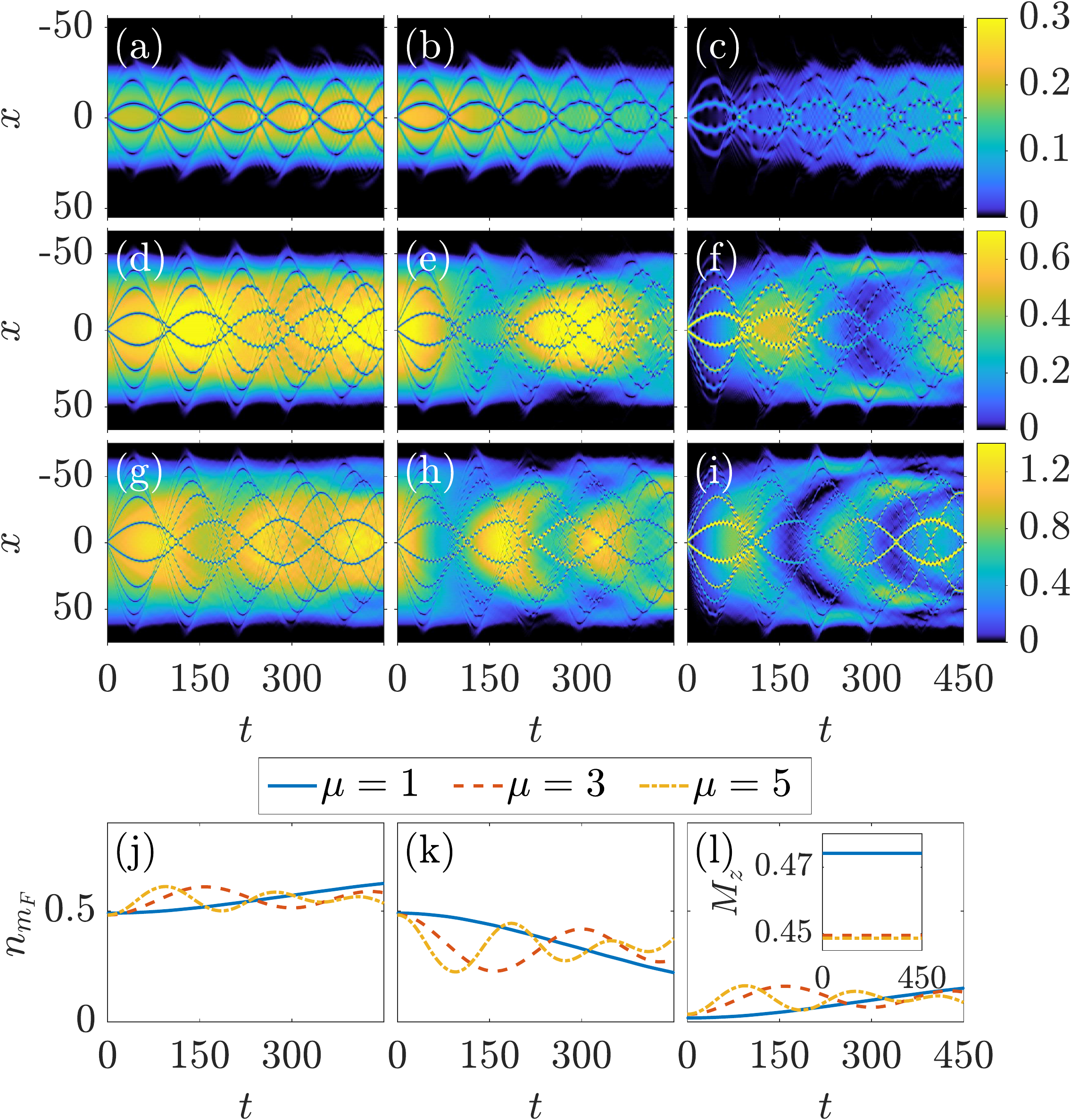}
\caption{Same as Fig.~\ref{fig:SDDB_trap} but upon varying the chemical potential $\mu$.
From top to bottom $\mu=1$, $\mu=3$ and $\mu=5$, allowing the generation of four [(a)-(c)], six [(d)-(f)] 
and eight [(g)-(i)] DDB solitons respectively.
In all cases, (a), (d), (g) [(b), (e), (h)] illustrate the formation of dark solitons in the $m_F={+1}$ [$m_F={0}$] component 
and (c), (f), (i) the generated bright solitons in the $m_F={-1}$ component of the spinor system.
Notice that the colormap has a $2.5:2:1$ ratio between the columns.
(j)-(l) Evolution of the normalized number of particles, $n_{m_F}(t)$, for each value of $\mu$.
The inset in (l) shows the total magnetization, $M_z(t)$, for each value of $\mu$. }
\label{fig:SDDB_trap_mu} 
\end{figure}

The above-discussed dynamical evolution of the spinor system holds equally
for all the different variations illustrated in Figs.~\ref{fig:SDDB_trap}(a)-(i).
However, the deformation of the DDB states is found to be delayed as $w^2$ is increased.
The latter result can be deduced by comparing at earlier evolution times
the density profile shown in Fig.~\ref{fig:SDDB_trap}(l) to the relevant ones illustrated in 
Figs.~\ref{fig:SDDB_trap}(j), (k).
Additionally, and also in all cases depicted in 
Figs.~\ref{fig:SDDB_trap}(a)-(i), 
the initially formed DDB structures that evolve later on into beating dark solitons
are seen to oscillate and interact within the parabolic trap.
However, while coherent oscillations are observed in Figs.~\ref{fig:SDDB_trap}(a), (d), (g),
incoherent ones occur when the number of states is increased (i.e. for increasing $w^2$).
In these latter cases, as shown in Figs.~\ref{fig:SDDB_trap}(b), (e), (h), 
several collision events between the outer and the inner beating states take place.
Despite the much more involved dynamical evolution of the spinor system in such cases, 
these beating states remain robust for all the evolution examples
that we have checked.  
Furthermore, we also explored the dynamical evolution of the spinorial BEC system 
for different values of the chemical potential, $\mu$.
Similarly to the aforementioned $w$ variation, 
a controlled formation of larger DDB arrays as $\mu$ increases can be once more verified.
The resulting states in increasing order, in terms of $\mu$, are presented in
Figs.~\ref{fig:SDDB_trap_mu}(a)-(i) for fixed $w^2=5$.    
Notice that since $w^2=5$ Figs.~\ref{fig:SDDB_trap_mu}(a)-(c) are respectively identical to 
Figs.~\ref{fig:SDDB_trap}(b), \ref{fig:SDDB_trap}(e) and \ref{fig:SDDB_trap}(h).
However, increasing $\mu$ increases the system's size. 
As such, arrays consisting of a larger number of DDB solitons are formed.
Indeed, six and eight DDB states are generated for $\mu=3$ and $\mu=5$ respectively. 
Importantly here, it is found that the presence of the spin interaction
has a more dramatic effect on the resulting states when compared to the previous variation.
Namely, the originally formed DDB structures transition into beating dark ones
much faster when compared to the $w$ variation. 
A case example can be seen in Fig.~\ref{fig:SDDB_trap_mu}(g), corresponding to $\mu=5$, 
where the dark solitons of the $m_F=+1$ component evolve into beating ones already at $t=150$.
Even for the largest $w^2=10$ value considered above, such a
transition occurs for this hyperfine 
component at evolution times $t \approx 300$ [see Fig.~\ref{fig:SDDB_trap}(c)]. 
To appreciate the effect of the spin interaction, we monitor 
during evolution the population, $n_{m_{F}}(t)$ ($m_F=+1,0,-1$),
of each hyperfine component and for all the different values of $\mu$
considered herein.
In particular, from left to right Figs.~\ref{fig:SDDB_trap_mu}(j)-(l) illustrate
$n_{+1}(t)$, $n_{0}(t)$ and $n_{-1}(t)$ respectively.
Notice that the population of each hyperfine component is affected more, the larger
the value of $\mu$ is. 
Evidently, the monotonic increase ($n_{\pm1}(t)$) 
or decrease ($n_0(t)$) for $\mu=1$ turns into damping oscillations as $\mu$
increases.
Such a coherent spin-mixing dynamics is in line with earlier predictions in 
spinor $F=1$ BECs~\cite{Pu1999,Chang2005}. 
Finally, we verified that the total magnetization $M_z(t)$ remains constant during evolution
acquiring a slightly smaller value as $\mu$ is increased [see the inset in Fig.~\ref{fig:SDDB_trap_mu}(l)].

\section{Conclusions and Future perspectives} \label{sec:summ_concl}

In this work the controlled creation of multiple soliton complexes 
of the DB type that appear in one-dimensional two-component, 
three-component and spinor BECs has been investigated.
Direct numerical simulations of each system's dynamical evolution have been performed    
both in the absence and in the presence of a parabolic trap.
In all models considered herein, the nucleation process is based
on the so-called matter wave interference of separated condensates
being utilized to study multi-component systems.
In this sense, this work offers a generalization of 
earlier findings in single-component setups 
to the much more involved multi-component ones, enabling the identification
of dark-bright solitons in two-component and dark-dark-bright and
dark-bright-bright solitons in three-component (and spinorial) gases. 
To achieve control over each system's dynamical evolution 
different parametric variations have been considered. 

In particular, for the homogeneous systems addressed in this effort,
inverse rectangular pulses were employed for the components
featuring interference,
and Gaussian ones for the remaining participating components.
Destructive interference of the two sides of the former pulse
leads to the nucleation of an array of dark soltions.
Additionally, the dispersion of the Gaussian pulse 
and its subsequent confinement in the effective potential 
created by each of the nucleated dark solitons 
results in the formation of bright solitons that are 
subsequently trapped and waveguided by their corresponding 
dark counterparts.  
It is found that manipulating the width of the IRP 
is sufficient to ensure the desired nucleation 
of multiple soliton compounds of the DB type.
This way, arrays of DB, and DDB and DBB solitons 
are dynamically produced in the two-component and spinor cases, respectively. 
Moreover, for the two-component system it is showcased that each of the 
generated DB solitons follows the analytical expressions
stemming from the integrable theory of the Manakov system.
The same holds true also for the DBB and DDB states nucleated in the three component system.
In the latter, generalized expressions that connect the soliton parameters are extracted and used 
to appreciate modifications of the soliton characteristics under different 
parametric variations.
While the same overall dynamical evolution is observed for the two- and three-component systems,
a significantly different picture can be drawn for the spinorial case.
Strikingly, and for a DDB nucleation process, it is found 
that during evolution the originally formed DDB soliton arrays 
begin to deform due to the spin-mixing dynamics.
The latter allows for exchange of particles between the hyperfine components. 
The aforementioned deformation leads in turn to the gradual formation
of arrays of beating dark states.
The latter, once formed, are seen to robustly propagate for large evolution
times. The existence of beating dark states in spinor systems
has not, to the best of our knowledge, been reported previously
and it is an interesting topic for further exploration.  

For the harmonically trapped scenarios our numerical findings suggest similar 
characteristics as in the homogeneous cases in terms of the nucleation
process, although naturally the dynamics is rendered more complex
due to the confinement and the induced interactions between the
produced solitary waves. 
In all cases it is found that by adjusting the width of the rectangular
pulse or the chemical potential of the participating 
components, the desirable number of DB, DDB and DBB soliton complexes can be generated. 
This provides a sense of dynamical control and design of
desired configurations in our system.
The number of the resulting coherent structures is found to increase upon increasing each of the above parameters. 
In the trapped case, the resulting multi-soliton arrays, irrespectively
of their type, are found to oscillate and interact
within the parabolic trap being robust for large evolution times.
Contrary to the above findings, for the spinorial BEC system 
a departure of the initially formed DDB states to the beating dark ones is 
showcased.  
Here, coherent spin-mixing dynamics is observed when monitoring 
the population of each hyperfine component. 
Damping oscillations of the latter occur, that are found to be enhanced upon 
increasing, for example, the chemical potential of each component.
Additionally, and also in comparison to the homogeneous case,
the beating dark states are formed faster in the trapped setting. 
This formation is further enhanced the larger the chemical
potential is. 
It is found that the beating dark solitons persist 
while oscillating and interacting with one another. 
The existence of these spinorial beating states can 
be tested in current state-of-the art experiments~\cite{Bersano2018},
and it is clearly a direction of interest in its own right for
future studies.
More specifically, it would be particularly interesting to generalize the
findings associated with the two-component beating dark solitons~\cite{Yan2012}
to the spinor case and study in a detailed manner the formation
and interactions of the spinor beating dark states identified herein.

Yet, another interesting perspective would be to compare and contrast the 
numerically identified DDB and DBB states of the three-component system 
to the analytical expressions that are available, at least for the integrable 
version of this model~\cite{biondini2016three}. More specifically,
one could generalize the criteria of the single component
IRP scenario obtained in the earlier works of~\cite{Zakharov1973}
to the formation of
both DB and also DDB or DBB solitons from similar initial
data in the multi-component case and compare
these predictions against the corresponding numerical computations.
Then, one could depart from the above Manakov limit and also study the fate of these structures
in non-integrable systems~\cite{katsimiga2018dark}, including the spinor one.
The breaking of integrability would allow in turn for effects such as the miscibility/immiscibility of the involved 
components to come into play~\cite{kiehn2019spontaneous}. 
The interplay of the resulting density variations with the potential persistence of
the solitary wave structures is an interesting topic for future study.
Also in the same context it would be interesting to systematically examine interactions 
between multiple DDB and DBB states. 
The role of other effects such as the potential Rabi coupling between the components could also
be of interest in its own right~\cite{pitaevskii2016bose,pitaevskii2016bose}.

Lastly, as has been discussed in relevant reviews such
as~\cite{Kevrekidis2016}, many of these ideas, such
as the DB solitons (generalizing to vortex-bright ones),
the beating dark solitons, etc., naturally generalize to corresponding
higher-dimensional states. Examining the potential for such states
as a result of interference or possibly other methods more concretely
associated with higher dimensions such as the transverse
instability would be of particular interest in its own right.

\vspace{0.5cm}
\section*{Acknowledgements}
 A.R.-R. thanks M. Pyzh and K. Keiler for helpful discussions.
This material is based upon work supported by the U.S.
\ National Science Foundation under Grant No.\ PHY-1602994 and under Grant No.\ DMS-1809074 (P.G.K.).
P.G.K. is also grateful to the Alexander von Humboldt Foundation for
support and to Mr. Kevin Geier
(University of Heidelberg) for preliminary iterations on the subject
of the present work. 


\bibliographystyle{apsrev4-1}
\bibliography{bib/SpontaneousGeneration}

\end{document}